\documentclass[a4paper,11pt]{article}
\usepackage{jheppub}
\usepackage[utf8]{inputenc}
\usepackage{float}
\usepackage{natbib}
\usepackage{graphicx}
\usepackage{comment}
\usepackage{wrapfig}
\usepackage{colortbl}
\usepackage{lineno}
\usepackage{microtype}
\usepackage[T1]{fontenc}
\usepackage{eurosym}
\usepackage{booktabs}
\usepackage{multirow}
\usepackage{amsmath}
\usepackage{todonotes}
\usepackage{lastpage}
\usepackage{enumitem}
\usepackage{lipsum}
\usepackage{physics}
\usepackage{longtable}
\usepackage{comment}
\usepackage{array}
\usepackage{caption}
\usepackage{subcaption}

\begin{document}

\definecolor{DeepPink}{rgb}{0.796,0.004,0.384}
\definecolor{RoyalBlue}{rgb}{0.02,0.016,0.667}
\definecolor{LavenderBlush}{rgb}{1.00,0.941,0.961}

\setlength{\columnseprule}{0.5pt}
\numberwithin{equation}{section}
\newcommand{\simas}[1]{\raisebox{-.1ex}{$\stackrel{\small{#1}}{\sim}$}}
\newcommand{\ssfill}{\xleaders\hbox to 0.35em{\scriptsize.}\hfill}
\makeatletter
\newcommand{\oset}[3][0ex]{\mathrel{\mathop{#3}\limits^{\vbox to#1{\kern-2\ex@\hbox{$\scriptstyle#2$}\vss}}}}
\makeatother
\newcommand*{\cventry}[7][.25em]{\noindent\begin{tabular*}{\textwidth}{l@{\extracolsep{\fill}}r}{\bfseries #4} & {\bfseries #5} \\{\itshape #3\ifthenelse{\equal{#6}{}}{}{, #6}} & {\itshape #2}\\\end{tabular*}\ifx&#7&\else{\\\vbox{\small#7}}\fi\par\addvspace{#1}}
\newcommand*{\hintfont}{\bfseries}
\newcommand*{\hintstyle}[1]{{\noindent\hintfont{#1}}}
\newcommand*{\cvitem}[3][.25em]{\ifthenelse{\equal{#2}{}}{}{\hintstyle{#2}: }{#3}\par\addvspace{#1}}
\let\OriginalQuotation\quotation
\renewcommand*{\quotation}{\OriginalQuotation\small\sf}
\newcommand{\tred}[1]{\textcolor{red}{#1}}
\newcommand{\tblue}[1]{\textcolor{blue}{#1}}
\newcommand{\be}{\begin{equation}}
\newcommand{\ee}{\end{equation}}
\newcommand{\ba}{\begin{array}}
\newcommand{\ea}{\end{array}}
\newcommand{\baa}{\begin{array}}
\newcommand{\eaa}{\end{array}}
\newcommand{\bea}{\begin{eqnarray}}
\newcommand{\eea}{\end{eqnarray}}
\newcommand{\half}{\frac{1}{2}}
\newcommand{\MS}{{\overline{\rm MS}}}
\newcommand{\LO}{{\rm LO}}
\newcommand{\NLO}{{\rm NLO}}
\newcommand\colorrow{\rowcolor{LavenderBlush}}

\title{Complex potential and the thermal width of quarkonium in quark-gluon plasma}

\author[a]{Rasmus Normann Larsen}
\author[b]{Peter Petreczky}
\author[b]{Jorge Luis Dasilva Golan}
\author[c,d]{Johannes~H.~Weber}


\affiliation[a]{Fakult\"at f\"ur Physik, Universit\"at Bielefeld, D-33615 Bielefeld, Germany}
\affiliation[b]{Physics Division, Brookhaven National Laboratory, Upton, New York 11973, USA}

\affiliation[c]{GSI Helmholtzzentrum f{\"u}r Schwerionenforschung GmbH, Planckstra{\ss}e~1, D-64291 Darmstadt, Germany}
\affiliation[d]{Institut f\"ur Kernphysik, Technische Universit\"at Darmstadt, Schlossgartenstra\ss e 2, D-64289 Darmstadt, Germany}

\abstract{
    We study the complex heavy quark anti-quark potential in lattice 2+1 flavor QCD at non-zero temperature. We point out a close relation between the imaginary part of the potential and the thermal quarkonium width. Using this relation, we estimate 
the thermal width of charmonia and bottomonia at high temperatures, which are not yet accessible with direct lattice QCD calculations of the quarkonium correlation functions and compare 
these to the results of the T-matrix analysis. For some of the excited quarkonia states, our estimates of the thermal width
are smaller than the ones obtained in the T-matrix analysis.
}

\maketitle

\section{Introduction}
\label{sec:intro}

It was first proposed by Matsui and Satz that quarkonium suppression can serve as a key signature of quark-gluon plasma (QGP) formation in heavy-ion collisions~\cite{Matsui:1986dk}. This idea is based on the expectation that color screening in a deconfined medium reduces the range of the interaction between a heavy quark and antiquark, ultimately leading to the dissolution of bound quarkonium states. Since then, the study of quarkonium properties in the QGP, as well as their production and suppression patterns in heavy-ion collisions, has developed into a major theoretical and experimental research program (see, e.g., Refs.~\cite{Aarts:2016hap,Zhao:2020jqu,Andronic:2024oxz} for recent reviews).

While color screening at high temperature is well established through lattice QCD calculations of the static quark-antiquark free energy~\cite{Bazavov:2018wmo}, its quantitative impact on quarkonium properties and the mechanism of quarkonium melting remain subjects of ongoing investigation. In the potential picture, the binding of a heavy quark-antiquark pair is described in terms of an effective interaction potential. This framework can be systematically justified within the effective field theory of potential non-relativistic QCD (pNRQCD)~\cite{Brambilla:1999xf,Brambilla:2004jw}, which has been extended to finite temperature~\cite{Laine:2006ns,Brambilla:2008cx}. At nonzero temperature, the potential becomes complex: in addition to modifications of its real part, it acquires an imaginary component that encodes medium-induced dissipation and decoherence effects~\cite{Laine:2006ns,Brambilla:2008cx}. Notably, the modification of the real part does not generally correspond to simple exponential (Debye) screening; for distances smaller than the Debye length, thermal effects lead instead to power-law modifications~\cite{Brambilla:2008cx}. In a weakly coupled QGP, the imaginary part of the potential plays a dominant role, leading to finite thermal widths and, ultimately, the dissolution of quarkonium states~\cite{Escobedo:2008sy,Laine:2008cf}.

Recent lattice QCD studies of quarkonium properties suggest that the dominant mechanism for quarkonium melting is the thermal broadening of bound states rather than color screening~\cite{Larsen:2019bwy,Larsen:2019zqv,Larsen:2020rjk,Ding:2025fvo,Larsen:2026qvs}. Specifically, these lattice QCD calculations find no significant changes in quarkonium masses within the estimated uncertainties, in contrast to the changes expected in the color-screening picture. On the other hand, all charmonium and bottomonium states studied acquire sizable thermal widths, which increase with temperature and are larger for more loosely bound states~\cite{Larsen:2019bwy,Larsen:2019zqv,Ding:2025fvo,Larsen:2026qvs}. Furthermore, a study of Bethe-Salpeter amplitudes at nonzero temperature does not indicate a change in the spatial structure of $S$-wave bottomonium states~\cite{Larsen:2020rjk}, as would be expected in potential models with a screened potential. Using general spectral decomposition of Wilson line correlators, Lattice QCD studies of the complex potential find no evidence for screening of the real part over the distances and temperatures studied, while the magnitude of its imaginary part increases with temperature and with the separation between the static quark and antiquark~\cite{Bala:2021fkm,Bazavov:2023dci}. This observation is consistent with the quarkonium properties discussed above. On the other hand, analyses of Wilson line correlators based on a hard thermal loop (HTL)-motivated spectral decomposition find instead a screened real part of the potential~\cite{Bala:2021fkm,Ali:2025iux}. The magnitude of the imaginary part nevertheless increases with temperature and separation in both approaches. Thus, we may expect that irrespective of the details of the spectral reconstruction of the Wilson line correlator the imaginary
part of the potential plays a role in quarkonium dissolution.

The thermal widths of quarkonium states play an important role in heavy-ion phenomenology. They provide essential input for the dissociation rates and, through detailed balance, the regeneration of quarkonium states in transport models
\cite{Zhao:2020jqu,Fu:2026ous,Wu:2026dmd}. Recent $T$-matrix calculations evaluate charmonium and bottomonium dissociation rates using in-medium interactions constrained by lattice QCD, including information from bottomonium correlators and Wilson line correlators~\cite{Wu:2025hlf}. While direct lattice QCD determinations of these widths would be preferable, the recent lattice calculations of bottomonium thermal widths with extended meson operators only cover temperatures up to 250~MeV. These calculations have been performed using non-relativistic QCD (NRQCD), and extending this approach to higher temperatures while maintaining sufficient temporal resolution would require smaller lattice spacings, eventually reaching the limit of applicability of lattice NRQCD~\cite{Ding:2025fvo}. On the other hand, while lattice QCD estimates of charmonium widths are already available up to 305~MeV~\cite{Larsen:2026qvs}, phenomenological applications require information about thermal widths at higher temperatures.

In this paper, we propose a relation between the imaginary part of the heavy quark-antiquark potential and the thermal widths of quarkonium states. We assess this relation through comparison with direct lattice QCD estimates of charmonium   \cite{Larsen:2026qvs} and bottomonium widths \cite{Ding:2025fvo}. This comparison reveals a close connection between the imaginary part of the potential and quarkonium thermal broadening. We then use the proposed relation to estimate widths at higher temperatures, beyond the range covered by these direct lattice calculations, and compare our findings with results from $T$-matrix calculations~\cite{Wu:2025hlf}. The rest of the paper is organized as follows. In Sec.~\ref{sec:complex}, we discuss the lattice determination of the complex potential, with particular emphasis on its imaginary part and finite volume effects in the determination
of the complex potential. In Sec.~\ref{sec:width}, we present our model relating the imaginary part of the potential to quarkonium thermal widths and compare the resulting estimates with lattice and $T$-matrix results. Our conclusions are presented in Sec.~\ref{sec:conclusions}.


\section{The complex potential at nonzero temperature}
\label{sec:complex}
 
The quark-antiquark ($Q \bar Q$) potential in QCD is defined as the energy of a static meson, in which the $Q$ and the $\bar Q$ are separated by a distance $r$. It can be obtained in lattice QCD by considering the decay of the Wilson loops or two Wilson lines in Coulomb gauge at large Euclidean time $\tau$ and spatial separation $r$. Examples include the zero-temperature determinations by RBC-Bielefeld~\cite{Cheng:2007jq}, MILC~\cite{Bernard:2000gd}, and HotQCD~\cite{Bazavov:2011nk}. To generalize the notion of the potential to nonzero temperature one needs to consider the spectral representation of the Wilson loop or the Wilson line correlator (WLC). The general spectral representation at temperature $T$ reads~\cite{Rothkopf:2011db}:
\begin{equation}
    W(\tau,r,T) = \int_{-\infty}^{+\infty} d\omega \,
    e^{-\omega\tau}\rho_r(\omega,T)\,,
    \label{eq:WLC}
\end{equation}
where $W(\tau,r,T)$ denotes the Wilson loop or Wilson line correlator, and the spectral function $\rho_r(\omega,T)$ in \eqref{eq:WLC} is labeled  by the distance $r$ between $Q$ and $\bar Q$. At zero temperature, the spectral function has the form
\begin{equation}
    \rho_r(\omega,0) = A_r\delta\left(\omega-V(r,0)\right) + \rho_r^{\rm high}(\omega)\,.
    \label{eq:spectral_T0}
\end{equation}
Here $V(r,0)$ is the zero-temperature static potential, $A_r$ denotes the overlap with the ground state, and $\rho_r^{\rm high}$ contains higher states, including excited states of $Q\bar Q$, as well as the scattering states containing hadrons and the static meson. Consequently, for sufficiently large Euclidean time the correlator is dominated by
\begin{equation}
    W(\tau,r,0) \simeq A_r e^{-V(r,0)\tau}.
    \label{eq:WLC_lowest_state}
\end{equation}

At finite temperature, interactions of the static pair with the thermal medium modify this spectral structure. If a well-defined dominant peak persists, its position characterizes the real part of the in-medium static energy, while the broadening of the peak encodes the dissipative part of the interaction. In this sense, the real and imaginary parts of the complex potential can be associated with the position and width, respectively, of the dominant structure in the static spectral function~\cite{Rothkopf:2011db}. This potential description requires the existence of such a dominant peak; at sufficiently high temperatures, where no well-defined peak remains, a potential interpretation might not be applicable. 

We revisit the determination of the complex potential in Ref.~\cite{Bazavov:2023dci} in two respects. First, we test whether the finite spatial volumes affect the medium-dependent contribution to the WLC. Second, we extend the cutoff of the Lorentzian peak from $2\Gamma$ to $4\Gamma$, motivated by the spectral shapes found in $T$-matrix calculations~\cite{Tang:2023tkm}.

The underlying ensembles (produced with an improved action~\cite{Weisz:1982zw,Luscher:1984xn}) contain $2+1$ flavors of HISQ{\color{blue}~\cite{Follana:2006rc}} quarks with a physical strange-quark mass. The lattice spacings $a=0.0493$, $0.0404$, and $0.0280$~fm correspond to $\beta=10/g_0^2=7.596$, $7.825$, and $8.249$, respectively, and cover temperatures $T=152$--$352$~MeV. At the two coarser spacings, we have $m_l=m_s/20$, corresponding to a pion mass of approximately $160$~MeV in the continuum limit. On the other hand, at the finest spacing we have $m_l=m_s/5$. These ensembles are used only for $T\ge195$~MeV, where the light-quark mass dependence is expected to be small~\cite{Bazavov:2023dci}.

We test the volume dependence of the two coarser lattice spacings using the ensembles summarized in Table~\ref{tab:pot_volumes}. The smaller finite-temperature ensembles have spatial extent $N_s=64$.  For the large-volume comparison, we use zero-temperature lattices of size $80^4$ and $96^4$ and finite-temperature lattices of size $96^3\times N_\tau$ and $128^3\times N_\tau$, respectively. The zero- and finite-temperature spatial sizes in each larger-volume pair are therefore different.
\begin{table}[tbp]
\centering
\begin{tabular}{cccccc}
\hline
$\beta$ & $a$ [fm] & $m_l/m_s$ & smaller $L$ [fm] & larger $T=0$ lattice & larger $T>0$ lattice \\
\hline
7.596 & 0.0493 & $1/20$ & 3.16 & $80^4$ & $96^3\times N_\tau$ \\
7.825 & 0.0404 & $1/20$ & 2.59 & $96^4$ & $128^3\times N_\tau$ \\
\hline
\end{tabular}
\caption{Ensembles used to test the volume dependence of the Wilson line correlator. The smaller spatial extent is $N_s=64$. The larger zero- and finite-temperature spatial sizes are $3.94$ and $4.73$~fm at $\beta=7.596$, and $3.88$ and $5.17$~fm at $\beta=7.825$.}
\label{tab:pot_volumes}
\end{table}

We reduce the statistical noise using gradient flow, which smooths the gauge fields over a radius $\sqrt{8\tau_F}$ at flow time $\tau_F$. We use the flow times of Ref.~\cite{Bazavov:2023dci} and fix the flowed configurations to Coulomb gauge with a stopping tolerance of $10^{-6}$. Further numerical details are given in the Appendix.

Our approach to obtaining the complex heavy-quark potential closely follows the analysis in Ref. \cite{Bazavov:2023dci}. For completeness, we first review it in some detail before proceeding to
the discussion of the numerical results. 

A convenient way to discuss and analyze the lattice QCD results on the WLC  is the effective mass,
\begin{equation}
    m_{\rm eff}(\tau,r,T) = -\partial_\tau\ln W(\tau,r,T)\,,
    \label{eq:meff_cont}
\end{equation}
which on the lattice is evaluated as
\begin{equation}
    m_{\rm eff}(\tau,r,T) = -\frac{1}{a} \ln\left[ \frac{W(\tau+a,r,T)} {W(\tau,r,T)} \right]\,.
    \label{eq:meff_latt}
\end{equation}
At zero temperature, suppression of the excited-state contributions with increasing $\tau$ leads to a plateau in this effective mass, whose value determines directly $V(r,0)$. At finite temperature the effective mass instead exhibits a nontrivial dependence on Euclidean time, reflecting the modified spectral structure of the static pair.
To determine the complex potential at nonzero temperature, we use the spectral ansatz 
\begin{equation}
    \rho_r(\omega,T) = \rho_r^{\rm low}(\omega,T) + \rho_r^{\rm peak}(\omega,T) + \rho_r^{\rm high}(\omega)\,.
    \label{eq:spec_split}
\end{equation}
The dominant structure $\rho_r^{\rm peak}$ contains the information associated with the complex potential, while $\rho_r^{\rm low}$ describes additional medium-dependent contributions at energies below the dominant peak. The high-energy contribution $\rho_r^{\rm high}$ is assumed to be insensitive to the temperature and can therefore be constrained from the zero-temperature correlator.

Using the zero-temperature decomposition of Eq. \eqref{eq:spectral_T0},
\begin{equation}
    W(\tau,r,0) = A_r\,e^{-V(r,0)\tau} + W^{\rm high}(\tau,r)\,,
    \label{eq:WLC_T0}
\end{equation}
the high-energy contribution can be determined as
\begin{equation}
    W^{\rm high}(\tau,r) = W(\tau,r,0) - A_r\,e^{-V(r,0)\tau}\,.
    \label{eq:WLC_high}
\end{equation}
In this way, the high-energy contribution can then be subtracted from the finite-temperature correlator in order to isolate the medium-dependent contributions.
For the high-energy contribution, we define the effective mass
\begin{equation}
    m_{\rm eff}^{\rm high}(\tau,r) = -\partial_\tau\ln W^{\rm high}(\tau,r)\,.
    \label{eq:meff_high}
\end{equation}

The parameters $A_r$ and $V(r,0)$ in Eq.~\eqref{eq:WLC_T0} are obtained from the ground-state region of the zero-temperature correlator. One can then define the subtracted finite-temperature correlator
\begin{equation}
    W_{\rm sub}(\tau,r,T) = W(\tau,r,T)-W^{\rm high}(\tau,r)\,.
    \label{eq:WLC_subtracted}
\end{equation}
The corresponding subtracted effective mass
\begin{equation}
    m_{\rm eff}^{\rm sub}(\tau,r,T) = -\partial_\tau\ln W_{\rm sub}(\tau,r,T)
    \label{eq:meff_subtracted}
\end{equation}
therefore provides enhanced sensitivity to the medium-dependent part of the static $Q\bar Q$ interaction. In particular, the behavior of $m_{\rm eff}^{\rm sub}$ at small Euclidean time is directly related to the cumulants of the subtracted spectral function, so the value of the effective mass near $\tau=0$ is governed by the first cumulant, while its initial slope is the negative of the second cumulant. A broadened dominant peak consequently produces a decreasing, approximately linear behavior of $m_{\rm eff}^{\rm sub}$ over a sufficiently small-$\tau$ interval.

\begin{figure}[t]
    \centering
    \includegraphics[page=4,width=0.45\linewidth]{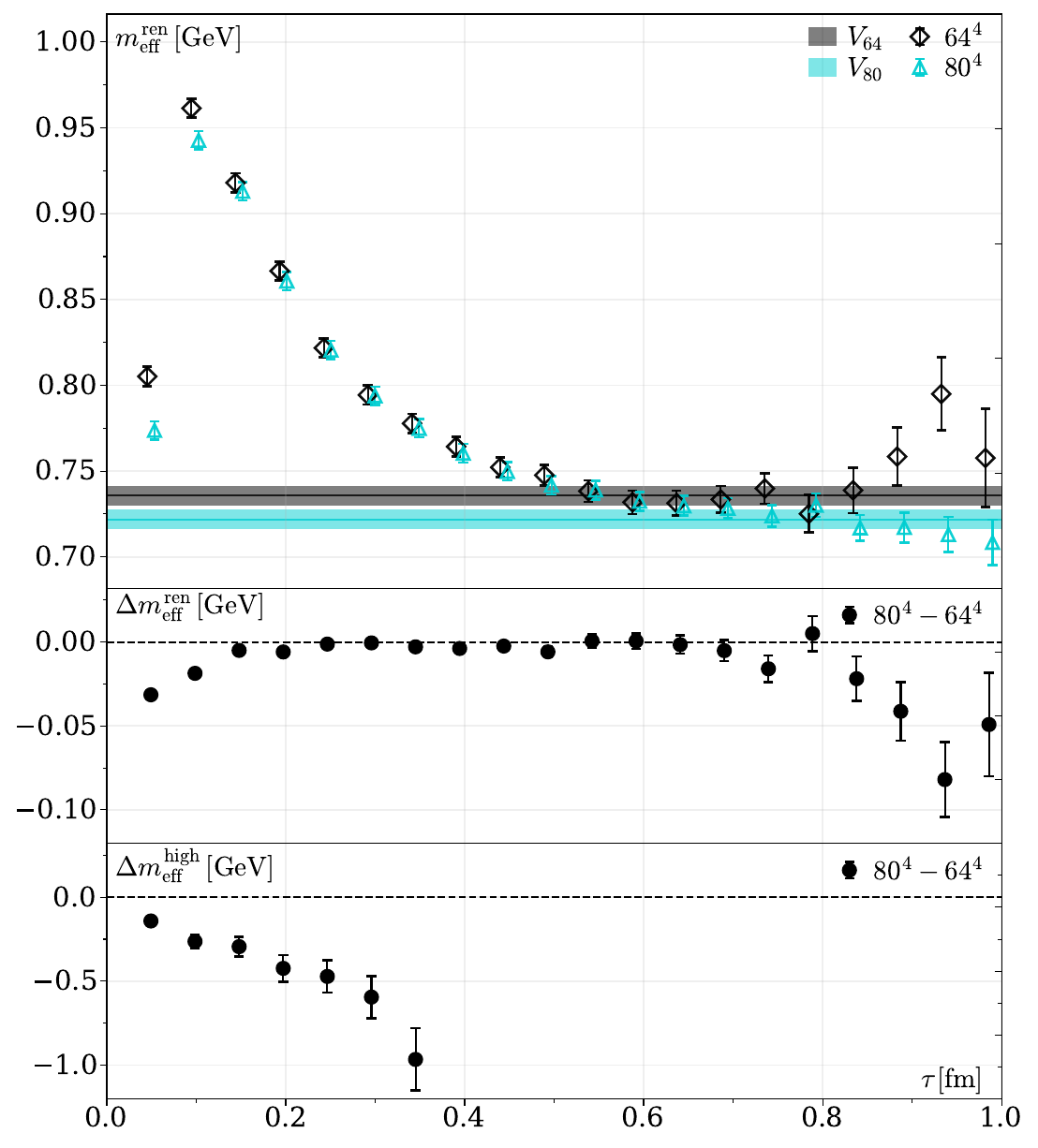}
    \caption{The effective mass of the Wilson line correlator at zero temperature for $\beta=7.596$ at a distance $r/a=15$ and $\tau_F/a^2=0.6$. The bands correspond to the ground state
    determination and its uncertainty (top panel). The middle panel shows the difference in the effective mass calculated on $80^4$ and $64^4$ lattices. The lowest panel shows the difference in $m_{\rm eff}^{\rm high}$ defined by Eq. (\ref{eq:meff_high}) calculated
    on $80^4$ and $64^4$ lattices. }
    \label{fig:meff_T0}
  \end{figure}

We first compare the zero-temperature effective masses on $64^4$ and $80^4$ lattices at $\beta=7.596$, $r/a=15$ ($r=0.739$~fm), and $\tau_F/a^2=0.6$. Figure~\ref{fig:meff_T0} shows a clear volume dependence at small Euclidean time and perhaps
less significant volume dependence at large Euclidean time. 
This suggests that the high energy contribution depends on the volume.
To study this effect in more detail, we calculate the effective mass
corresponding to the high energy part of the correlation function,
$m_{\rm eff}^{\rm high}$.
The difference in $m_{\rm eff}^{\rm high}$ (shown in the lower panel of Fig.~\ref{fig:meff_T0}) is larger at the smallest times than the difference in the full effective mass, clearly showing large finite volume effect
in the high energy part of the spectral function. The figure also indicates that there is a slight downward shift
in the ground state when we perform the calculation at a larger volume. Similar analysis was performed for $\beta=7.825$
as shown in the Appendix \ref{sec:setup}.
In Fig. \ref{fig:potT0} we show
the zero temperature potential as a function of $r$ calculated with different volumes for $\beta=7.596$ and $7.825$.
In both cases we see small but visible volume dependence for $r>0.7$ fm. The effect is larger for $\beta=7.825${\color{blue},} since the difference in the physical volume is bigger in this case. 
Since the static potentials obtained on different spatial volumes agree at small separations, it is unlikely that the renormalization condition depends on the volume.

\begin{figure}[t]
    \centering
    \includegraphics[width=0.45\linewidth]{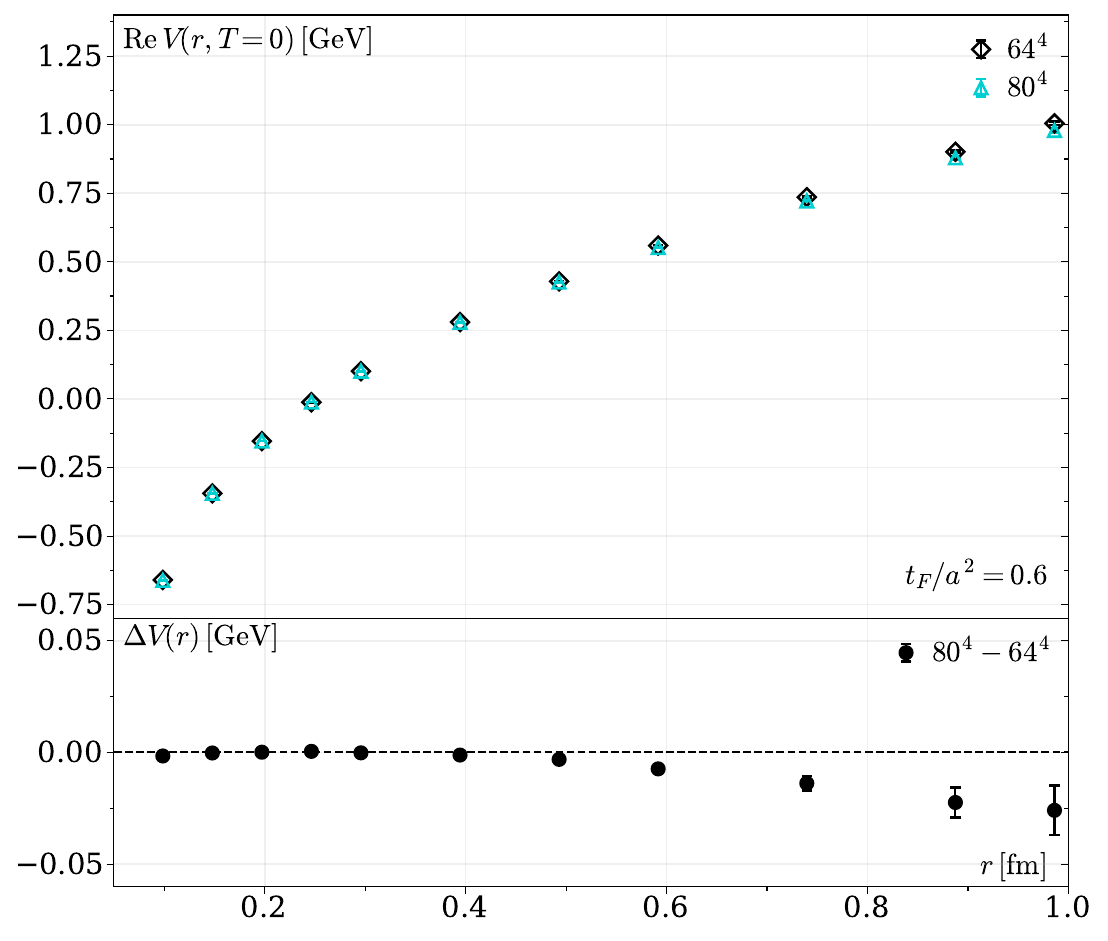}
    \includegraphics[width=0.45\linewidth]{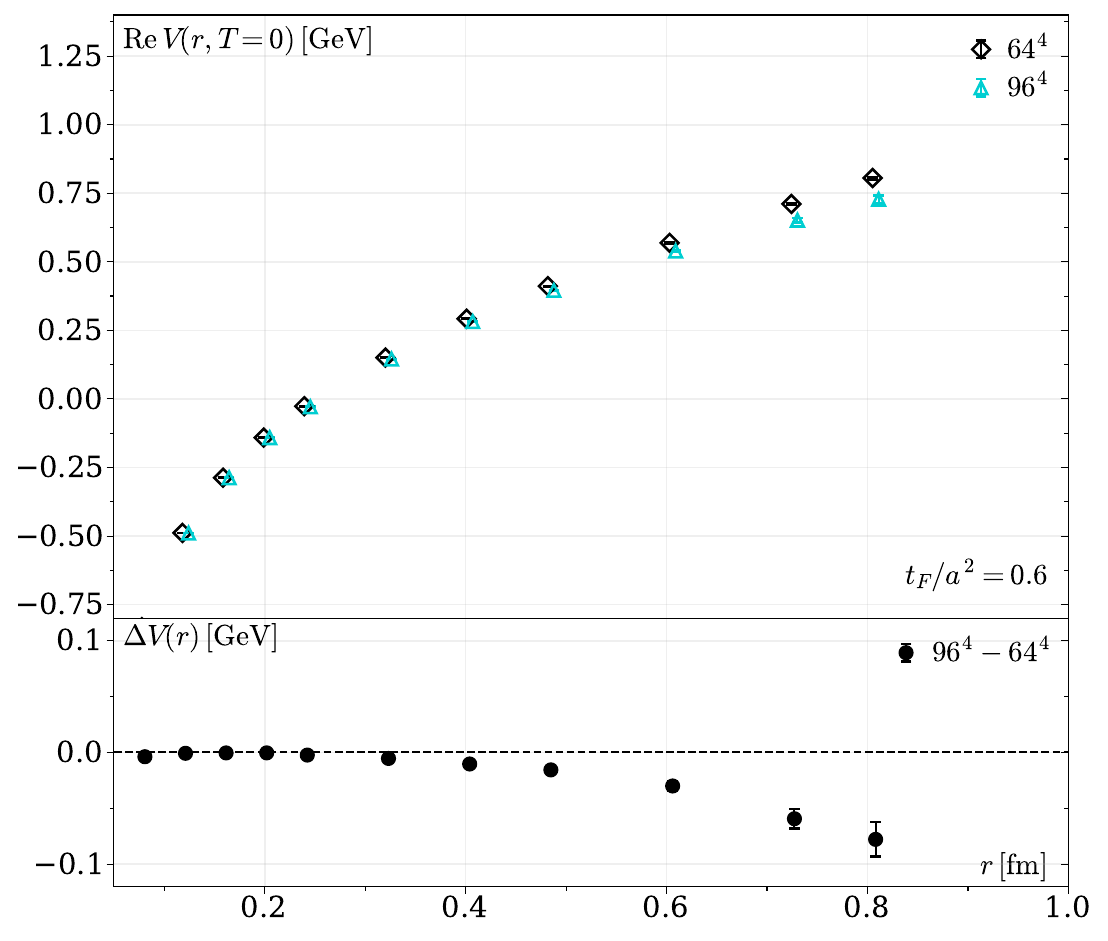}
    \caption{Finite-volume comparison of the zero-temperature static potential $\rm{Re}V(r,0)$ as a function of the quark-antiquark separation $r$, at flow time $\tau_F/a^2=0.6$. The left panel shows the results obtained at $\beta=7.596$, while the right panel compares lattices at $\beta=7.825$.}
    \label{fig:potT0}
 \end{figure}
 
\begin{figure}[t]
    \centering
    \includegraphics[page=4,width=0.45\linewidth]{pictures/l6496_Nt16_b7596_direct_meff_r15.pdf}
    \includegraphics[width=0.45\linewidth]{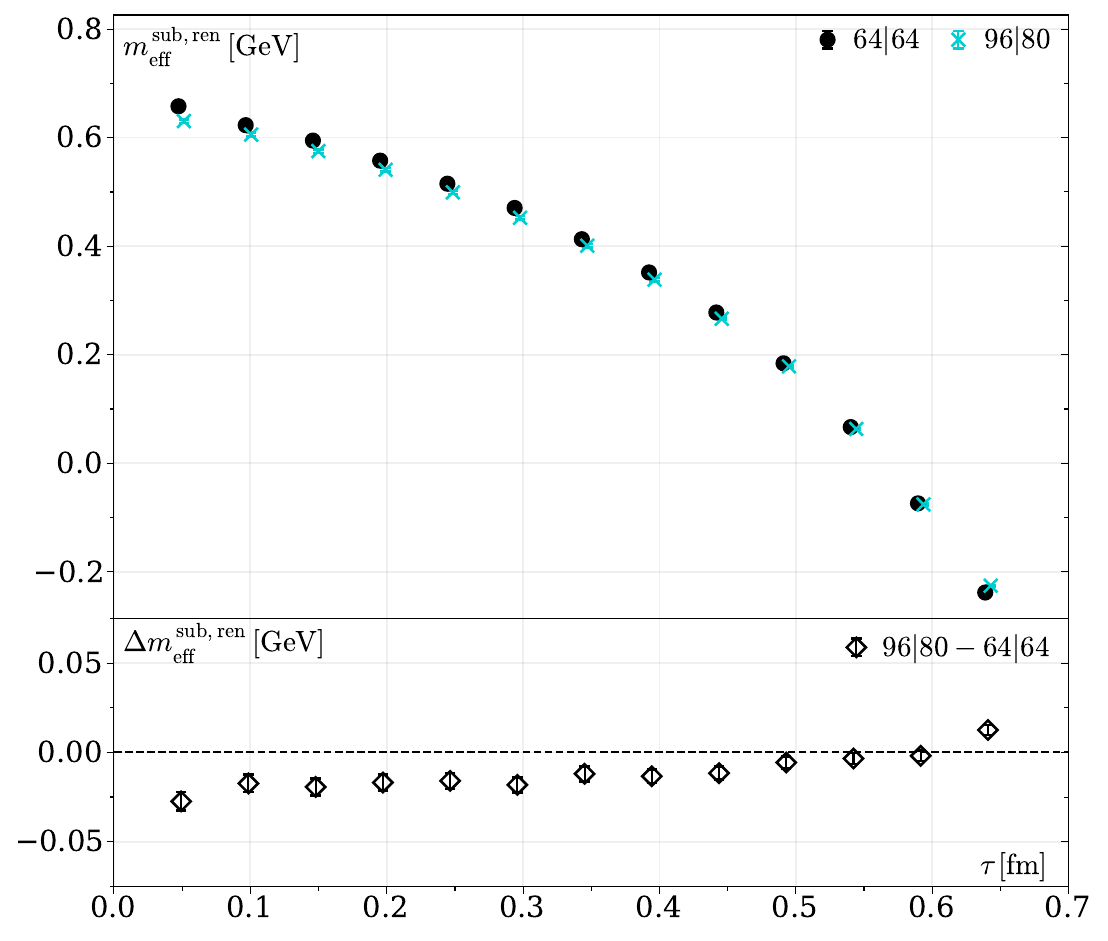}
    \caption{Volume dependence of effective masses at $T=250$~MeV, $\beta=7.596$, $r/a=15$ ($r=0.739$~fm), and $\tau_F/a^2=0.6$. Left: un-subtracted effective masses on $64^3\times16$ and $96^3\times16$ lattices and their difference. Right: subtracted effective masses for the $64|64$ and $96|80$ combinations and their difference. The notation $N_s^T|N_s^0$ specifies the thermal and vacuum spatial extents used in the subtraction.}
    \label{fig:meff_T}
 \end{figure}

At finite temperature, we compare $64^3\times16$ and $96^3\times16$ lattices at the same coupling, separation, and flow time, corresponding to $T=250$~MeV. The un-subtracted effective masses in the left panel of Fig.~\ref{fig:meff_T} show a volume dependence at small Euclidean time similar to that at zero temperature. There is also some volume dependence at the largest value of the Euclidean time available in our study. This can be attributed to the volume dependence of the spectral function at very small energies.

For the subtraction, we pair the $64^3\times16$ correlator with the $64^4$ vacuum correlator and the $96^3\times16$ correlator with the $80^4$ vacuum correlator. We denote these combinations by $64|64$ and $96|80$, where the first number specifies the thermal spatial extent and the second the vacuum spatial extent in lattice units. Overall we see the reduction of the finite volume effects in the subtracted effective masses. 

Note that there is a difference in the volume used at zero and finite temperature calculations for the subtraction. This might be the reason why we see the largest finite volume effect at the smallest $\tau$. At intermediate $\tau$ we see a small downward shift consistent with the same shift seen at $T=0$. Finally, at larger $\tau$ the volume effects are smaller due to the contribution from the spectral function at very small energies. We performed a similar analysis of the subtracted and un-subtracted effective masses at other temperature values, as shown in Appendix \ref{sec:setup}.

We extract the complex potential from the Wilson line correlators of Ref.~\cite{Bazavov:2023dci} using an explicit parametrization of the medium-dependent spectral contributions:
\begin{align}
\rho_r^{\rm peak}(\omega,T)
 &=\frac{A_r^{\rm peak}(T)}{\pi}
 \frac{\Gamma(r,T)}{[\omega-V_{\rm R}(r,T)]^2+\Gamma^2(r,T)}
 \Theta\!\left(4\Gamma(r,T)-|\omega-V_{\rm R}(r,T)|\right),
 \label{eq:pot_peak}\\
\rho_r^{\rm low}(\omega,T)
 &=A_r^{\rm low}(T)\,\delta\!\left(\omega-\omega_r^{\rm low}(T)\right).
 \label{eq:pot_low}
\end{align}
Here $V_{\rm R}$ is the peak position, $\Gamma$ is its Lorentzian half-width, and $\Theta$ is the step function. The amplitudes $A_r^{\rm peak}$ and $A_r^{\rm low}$ specify the strengths of the two contributions, while $\omega_r^{\rm low}$ locates the additional low-energy contribution. We use ${\rm Im}\,V=\Gamma>0$ to denote the positive broadening parameter; the corresponding dissipative potential is $V_{\rm R}-i\Gamma$, consistently with the convention in Sec.~\ref{sec:width}.

The cutoff in Eq.~\eqref{eq:pot_peak} is $4\Gamma$, compared with $2\Gamma$ in Ref.~\cite{Bazavov:2023dci}. This choice is motivated by $T$-matrix calculations of the infinitely heavy $Q\bar Q$ spectral function, which find a Lorentzian shape over an energy interval of at least $\pm4\Gamma$ around the peak~\cite{Tang:2023tkm}. Increasing the cutoff gives more weight to the spectral tails and increases the spectral spread at fixed $\Gamma$. A smaller peak width can therefore reproduce the same effective-mass slope. This explains why the revised imaginary part of the potential is smaller than the earlier estimate, and connects its extraction to the spectral-shape dependence of the quarkonium widths discussed in Sec.~3.

\begin{figure}[t]
    \centering
    \includegraphics[width=0.48\linewidth]{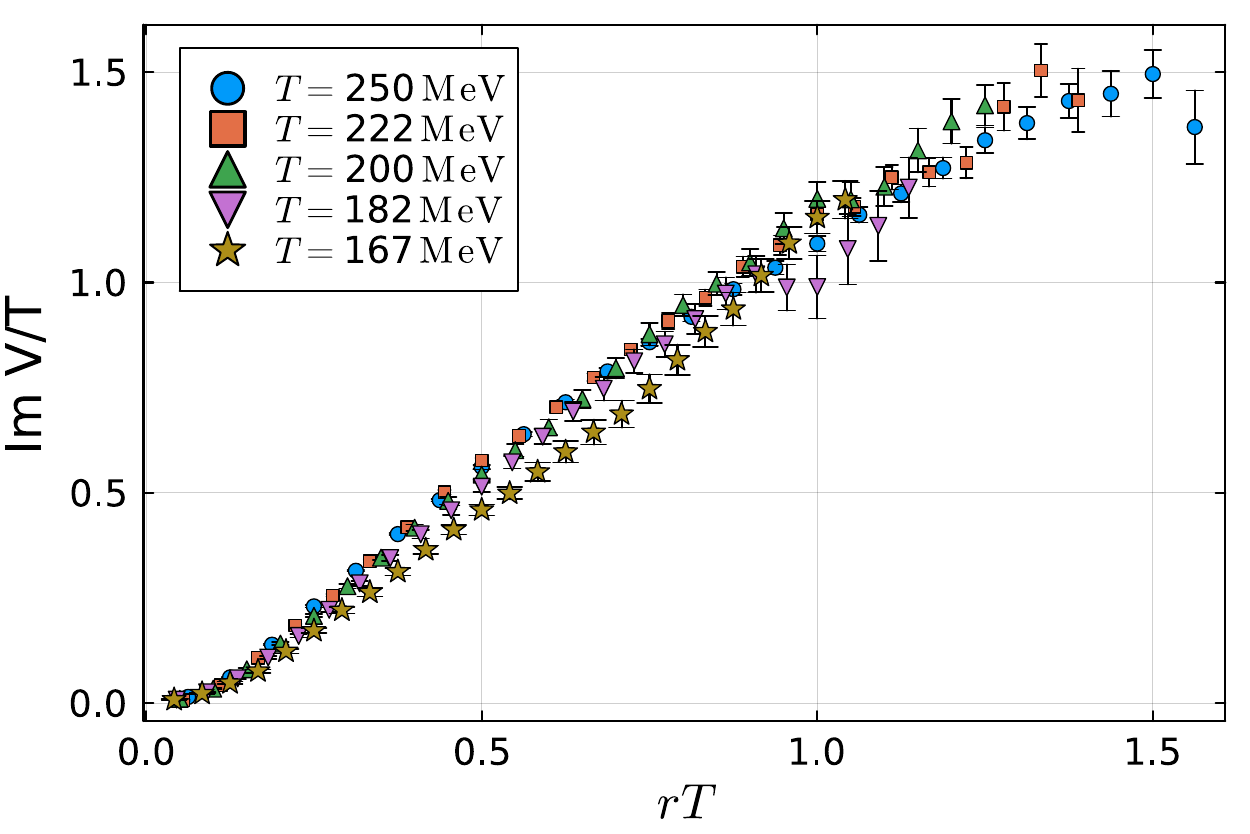}
    \includegraphics[width=0.48\linewidth]{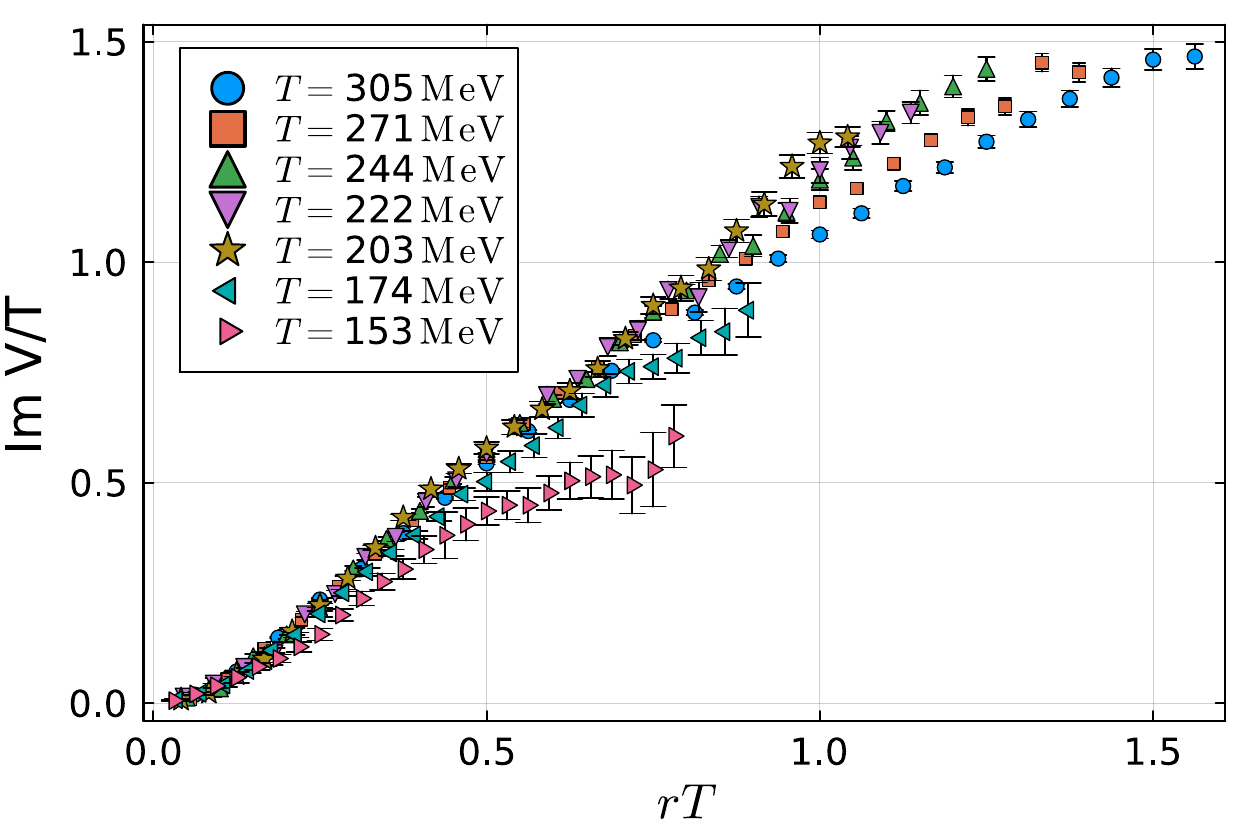}
     \includegraphics[width=0.48\linewidth]{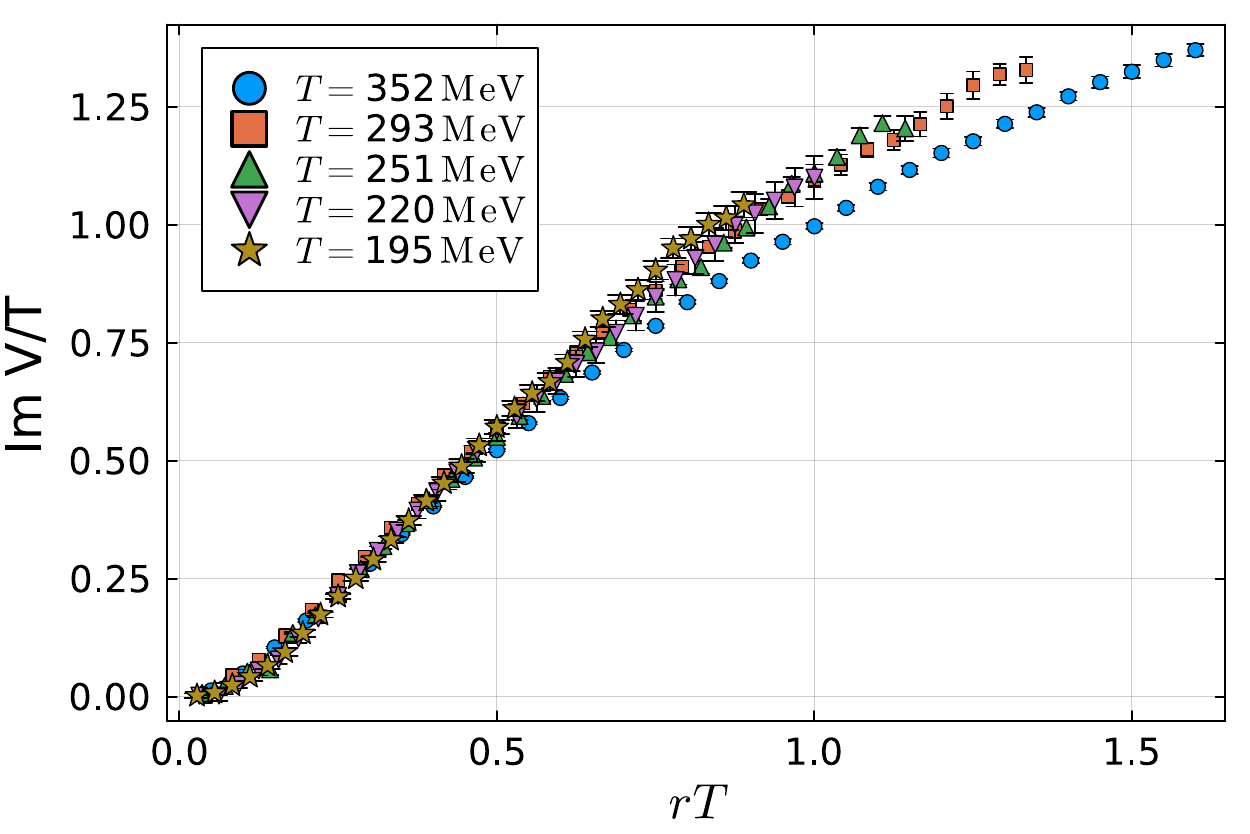}
     \includegraphics[width=0.48\linewidth]{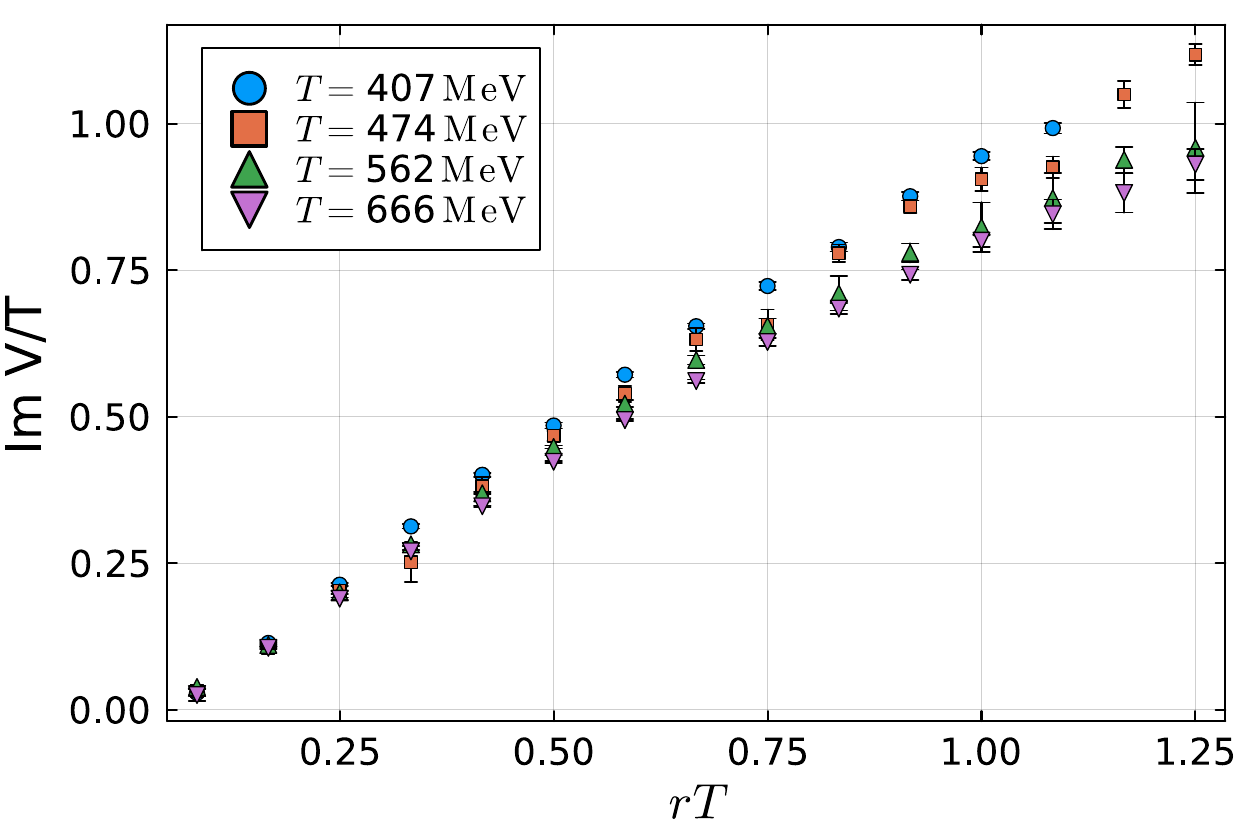}
\caption{The imaginary part of the potential as a function of $r$ at different temperatures  calculated for $\beta=7.596$ (top left),  $\beta=7.825$ (top right), $\beta=8.249$ (bottom left) and 4 different $\beta$ values with a smaller volume of $48^3$(bottom right).}
\label{fig:ImV}
\end{figure}

The fitted peak position is temperature independent within uncertainties and agrees with the zero-temperature potential over the distances and temperatures studied. The positive imaginary part is shown in Fig.~\ref{fig:ImV}. It increases with separation and temperature, indicating stronger broadening for more widely separated static pairs at higher temperatures. 
These results provide the input for the wave-function averages in Sec.~\ref{sec:width}, where this separation dependence generates the hierarchy of quarkonium widths.
Fig.~\ref{fig:ImV} suggests approximate temperature scaling for $T\approx 180$ to $300$ MeV. At higher temperatures, there are hints that the positive imaginary part is slightly lower than temperature scaling. 
We include in the lower right panel results at higher temperatures that support this trend, which have been obtained from a reanalysis of WLC data on $48^3\times 12$ lattices~\cite{Bala:2021fkm} using Eq.~\eqref{eq:pot_peak}. Such behavior is expected for high temperatures from the presence of powers of the temperature dependent coupling in the leading order imaginary part~\cite{Laine:2006ns,Brambilla:2008cx}.

We also performed the spectral analysis using Eqs. (\ref{eq:spec_split}), (\ref{eq:pot_peak}) and (\ref{eq:pot_low}) of the WLC 
obtained in the large volumes. We find that the volume dependence of the effective masses discussed above are mostly absorbed 
in the high and low energy part of the spectral functions and the parameters entering $\rho_r^{\rm peak}(\omega,T)$
for the $r$-range shown in Fig.~\ref{fig:ImV}. Thus, the use of the imaginary part of the potential obtained in the small volume is justified.

Finally, we mention that an alternative analysis employs a hard thermal loop (HTL)-motivated spectral representation for the logarithm of the Wilson correlator~\cite{Bala:2019cqu,Bala:2021fkm,Ali:2025iux}. This approach yields a screened real potential close to the color-singlet free energy in Coulomb gauge. Although it provides a practical means of extracting a complex potential, its theoretical justification in the temperature range considered here remains unclear. In particular, the separation between the thermal scale and the softer screening scale underlying the leading-order HTL approximation is not established, so corrections to the assumed spectral structure need not be small. The inability of the leading-order HTL prediction to reproduce the Euclidean-time dependence of the lattice WLC~\cite{Bala:2021fkm} reinforces this concern. Thus, the screened real potential obtained in this approach should be understood in conjunction with the additional spectral assumptions entering the extraction.


\section{Complex potential and quarkonium widths}
\label{sec:width}

As discussed in Sec.~\ref{sec:intro}, recent lattice QCD studies find no significant shifts in quarkonium masses within the estimated uncertainties, while the states acquire sizable thermal widths~\cite{Ding:2025fvo,Larsen:2026qvs}. The absence of a mass shift alone does not establish that the internal structure remains unchanged. Additional support for this picture comes from the study of bottomonium Bethe-Salpeter amplitudes in Ref.~\cite{Larsen:2020rjk}: the dependence of the corresponding correlators on the quark-antiquark separation shows little temperature dependence for the $S$-wave states studied (see Figs.~9 and 10 of that reference). These observations motivate an approximation in which the spatial wave functions retain their vacuum form, while the interaction with the medium generates a temperature-dependent width. 
A similar pattern is found for a static $Q\bar Q$ pair. In the analysis based on the general spectral decomposition discussed in Sec.~\ref{sec:complex}, the real part of the potential remains approximately unchanged over the distances and temperatures studied, whereas the magnitude of its imaginary part increases with temperature and separation~\cite{Bazavov:2023dci}. 
Furthermore, the lattice results show no significant dependence of the thermal width on the heavy-quark spin orientation: the widths of $\eta_c$ and $J/\psi$, as well as those of $\chi_{c0}$ and $\chi_{c1}$, agree within the estimated uncertainties~\cite{Larsen:2026qvs}. Similarly, in the bottomonium sector, the widths agree within uncertainties for the spin partners $\eta_b$ and $\Upsilon$, and for $\chi_{b0}$, $\chi_{b1}$, and $h_b$ with the same radial quantum number.
This suggests a simple connection between the thermal broadening of a static pair and that of a physical quarkonium state: average the broadening of the static pair over the distribution of quark-antiquark separations in the bound state.

We therefore estimate the width of a quarkonium state $\alpha$ as
\begin{equation}
\Gamma_{\alpha}(T)=\int d^3x\, |\psi_{\alpha}(\mathbf{x})|^2\, {\rm Im}\,V(r,T),\qquad r=|\mathbf{x}|.
\label{eq:width}
\end{equation}
The spatial wave function $\psi_{\alpha}(\mathbf{x})$ is normalized according to
\begin{equation}
\int d^3x\,|\psi_{\alpha}(\mathbf{x})|^2=1.
\label{eq:wave_norm}
\end{equation}
We use ${\rm Im}\,V$ to denote the positive broadening parameter, so that the dissipative potential entering real-time evolution is $V_{\rm R}-i\,{\rm Im}\,V$. The quantity $\Gamma_{\alpha}$ denotes the Lorentzian half-width parameter used in the lattice quarkonium analyses~\cite{Ding:2025fvo,Larsen:2026qvs}. For an isolated Lorentzian peak, its full width at half maximum is $2\Gamma_{\alpha}$; a pole at $E_{\alpha}-i\Gamma_{\alpha}$ gives a probability decay rate $2\Gamma_{\alpha}$. These conventions must be kept consistent when comparing spectral widths with reaction rates.

Equation~(\ref{eq:width}) assumes that a quarkonium configuration with separation $r$ experiences the same local broadening as a static pair at that separation. Averaging over the wave function then gives the width of the physical state. The same expression follows from first-order perturbation theory for the imaginary energy shift generated by $-i\,{\rm Im}\,V$. Its use beyond the perturbative regime is a model assumption, which we assess by comparison with direct lattice estimates. In particular, the approximation neglects medium-induced changes in the wave functions and explicit finite-mass corrections to the interaction with the medium. It also provides a natural explanation for the hierarchy of widths: states with greater weight at large separations sample a larger imaginary potential.

The Bethe-Salpeter amplitudes obtained in Ref.~\cite{Larsen:2020rjk} provide useful guidance for choosing the wave functions. That analysis shows that the amplitudes of $\Upsilon(1S)$, $\Upsilon(2S)$, and $\Upsilon(3S)$ can be reproduced well by solutions of the Schr\"odinger equation with the zero-temperature lattice potential and a bottom-quark mass parameter of about $5.6$~GeV. This lattice-motivated bottom-quark mass is considerably larger than those used in most potential models summarized in Ref.~\cite{QuarkoniumWorkingGroup:2004kpm}, but only slightly larger than the value $m_b=5.18$~GeV used in the Cornell potential model~\cite{Eichten:1979ms}. We therefore calculate the $1S$, $2S$, and $3S$ bottomonium wave functions using a parametrization of the zero-temperature lattice potential and vary the mass parameter over $m_b=5.0$--$5.8$~GeV. 

Using these wave functions and the imaginary potential determined in Sec.~\ref{sec:complex}, we estimate the widths from Eq.~(\ref{eq:width}). In the numerical integration, we linearly interpolate ${\rm Im}\,V$ between the available lattice separations. Beyond the largest available separation, we hold ${\rm Im}\,V$ fixed at its value at that separation, or assume it to rise linearly, and use the two limiting cases to estimate the associated systematic uncertainty of the result. We propagate the uncertainties in ${\rm Im}\,V$ and estimate the mass-parameter dependence from the spread of the results over $m_b=5.0$--$5.8$~GeV. These two contributions are added in quadrature.

\begin{figure}[tbp]
\centering
\includegraphics[width=0.48\textwidth]{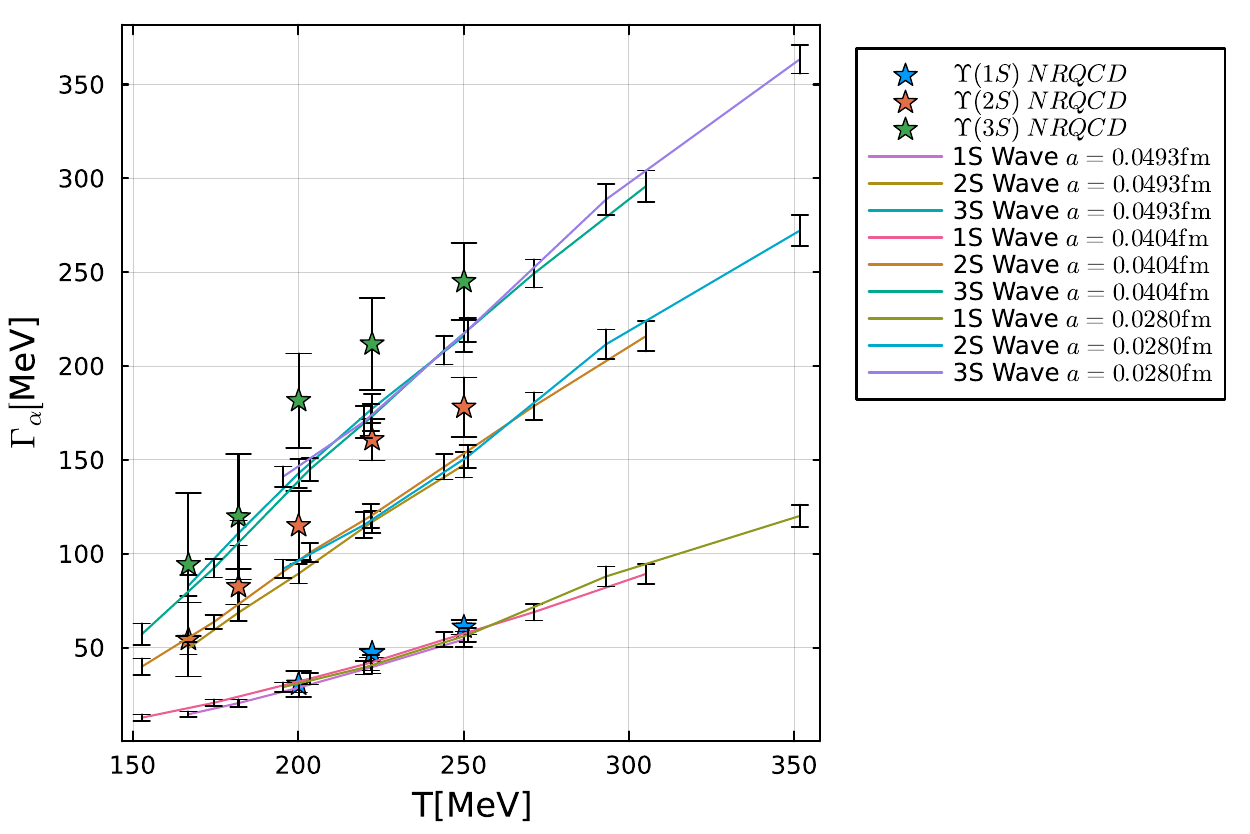}
\includegraphics[width=0.48\textwidth]{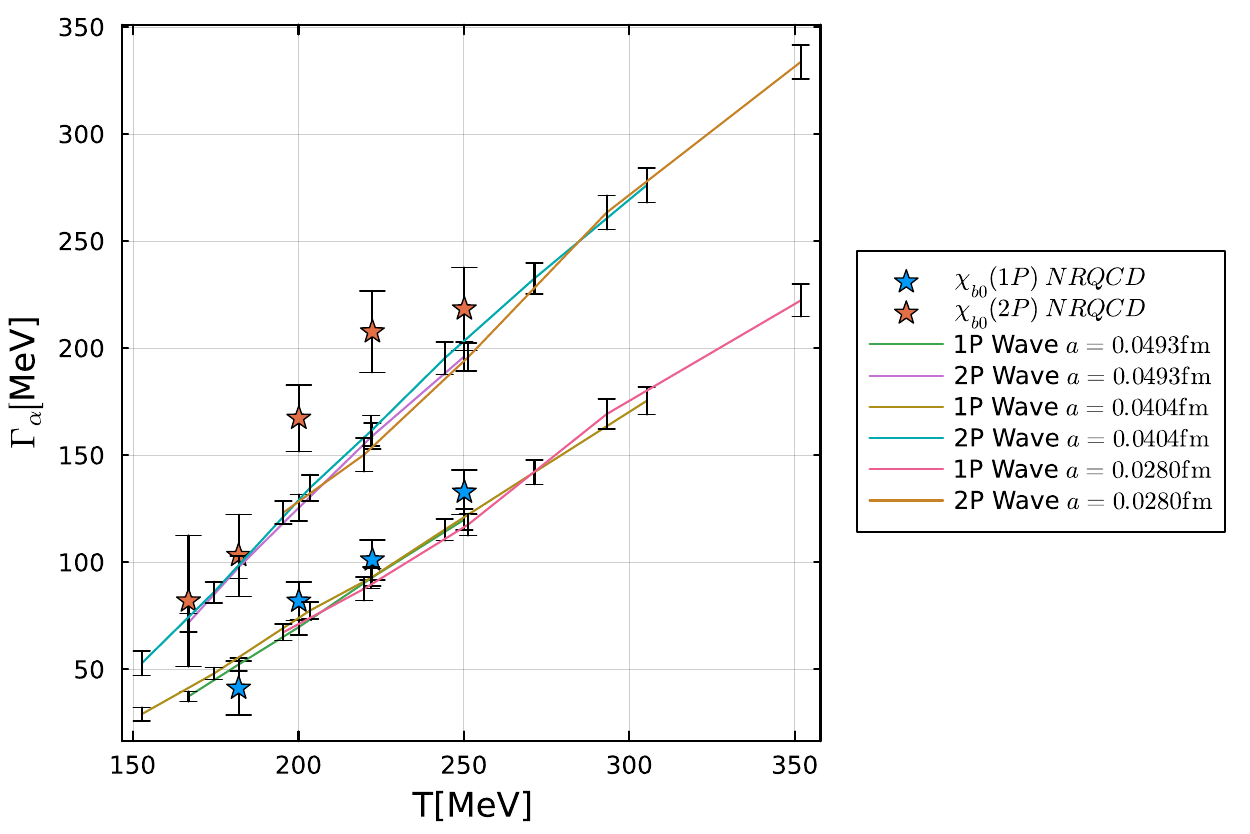}
\caption{Thermal widths of $\Upsilon(1S)$, $\Upsilon(2S)$, and $\Upsilon(3S)$ (left) and $\chi_{b0}(1P)$ and $\chi_{b0}(2P)$ (right). The estimates from Eq.~(\ref{eq:width}), shown by the smaller symbols connected by lines, are compared with direct lattice NRQCD results~\cite{Ding:2025fvo}, shown by the larger symbols. The estimates use wave functions obtained from the zero-temperature lattice potential, with the choice of bottom-quark mass guided by the Bethe-Salpeter analysis of Ref.~\cite{Larsen:2020rjk}. Widths are given in the half-width convention described in the text.}
\label{fig:width_bot}
\end{figure}

The resulting widths are compared with direct lattice NRQCD estimates~\cite{Ding:2025fvo} in the left panel of Fig.~\ref{fig:width_bot}. 
The model describes the $\Upsilon(1S)$ width well throughout the temperature range shown. For the excited states, it reproduces the ordering and overall increase with temperature, but generally gives smaller widths. The $\Upsilon(2S)$ estimates agree with the lattice results at the lower temperatures and fall below them around and above $T=200$~MeV. The $\Upsilon(3S)$ estimates are broadly compatible with the lattice results within their larger uncertainties. The wave functions and thermal widths of the $1P$ and $2P$ bottomonium states are obtained in the same way. The resulting widths are compared with the lattice NRQCD results for $\chi_{b0}(1P)$ and $\chi_{b0}(2P)$ in the right panel of Fig.~\ref{fig:width_bot}. The agreement is good for the $1P$ state. For the $2P$ state, the model captures the temperature dependence but underestimates the central lattice values, most visibly around $T=200$--$220$~MeV. Taken together, these comparisons support the proposed relation as an approximate description of bottomonium thermal broadening, while also showing state-dependent deviations.

\begin{figure}[tbp]
\centering
\includegraphics[width=0.48\textwidth]{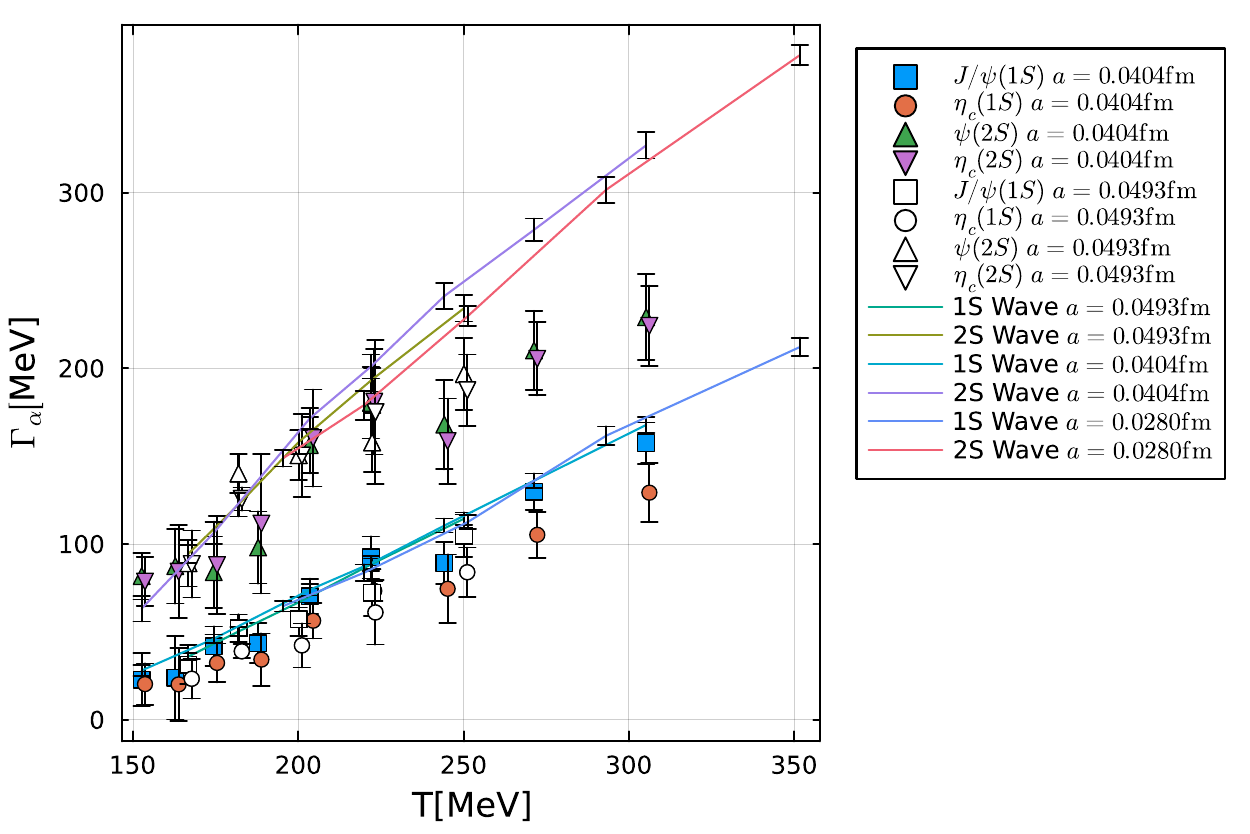}
\includegraphics[width=0.48\textwidth]{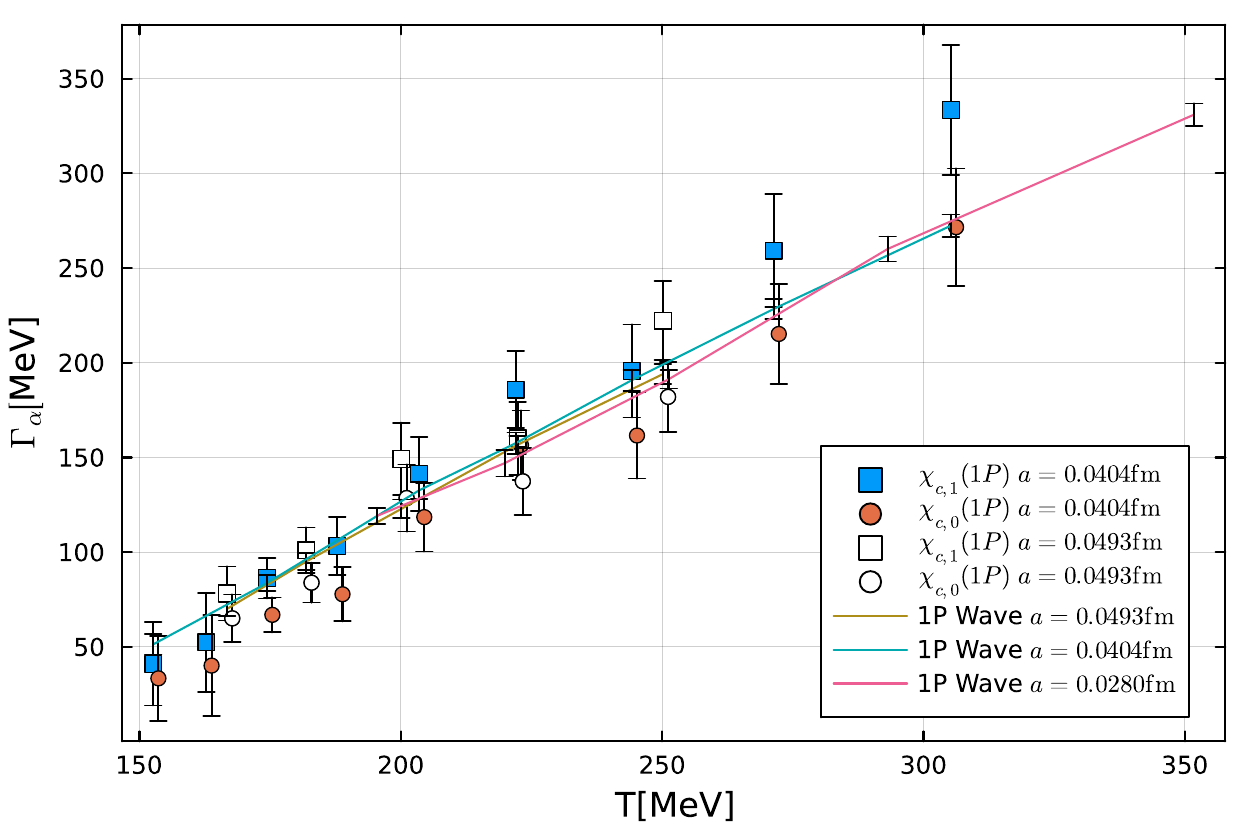}
\caption{Thermal widths of $1S$ and $2S$ charmonia (left) and $1P$ charmonia (right). Lines show the estimates from Eq.~(\ref{eq:width}); symbols show lattice results from Ref.~\cite{Larsen:2026qvs}. Filled and open symbols correspond to lattice spacings $a=0.0404$ and $0.0493$~fm, respectively. The same model estimate is used for the two spin channels within each radial and orbital multiplet. Widths are given in the half-width convention described in the text.}
\label{fig:width_charm}
\end{figure}

We next consider charmonium, comparing our estimates with recent lattice results obtained using a HYP-smeared clover action~\cite{Larsen:2026qvs}. Motivated by our experience in the bottomonium sector, we calculate the $1S$, $2S$, and $1P$ wave functions using charm-quark mass parameters in the range $m_c=1.8$--$2.0$~GeV. These values are larger than those used in most potential models summarized in Ref.~\cite{QuarkoniumWorkingGroup:2004kpm}, while the Cornell-model value $m_c=1.84$~GeV lies within our chosen range~\cite{Eichten:1979ms}.
 We therefore compare the $1S$ estimate with both $\eta_c$ and $J/\psi$, the $2S$ estimate with $\eta_c(2S)$ and $\psi(2S)$, and the $1P$ estimate with $\chi_{c0}$ and $\chi_{c1}$.
The comparison is shown in Fig.~\ref{fig:width_charm}. The model reproduces the overall temperature dependence and the hierarchy between the $1S$, $1P$, and $2S$ widths. The $1S$ and $1P$ estimates give a reasonable description of the lattice results, with a common spin-independent estimate for each multiplet. For the $2S$ states, the model increasingly overestimates the lattice widths above approximately $225$~MeV, reaching a discrepancy of {\color{blue} about} 
a factor of $1.5$ at some temperatures. Thus the simple spatial average captures the principal trends for both heavy-quark flavors, although some details of the temperature dependence of the excited-state widths are not reproduced. 

\begin{figure}[tbp]
\centering
\includegraphics[width=0.48\textwidth]{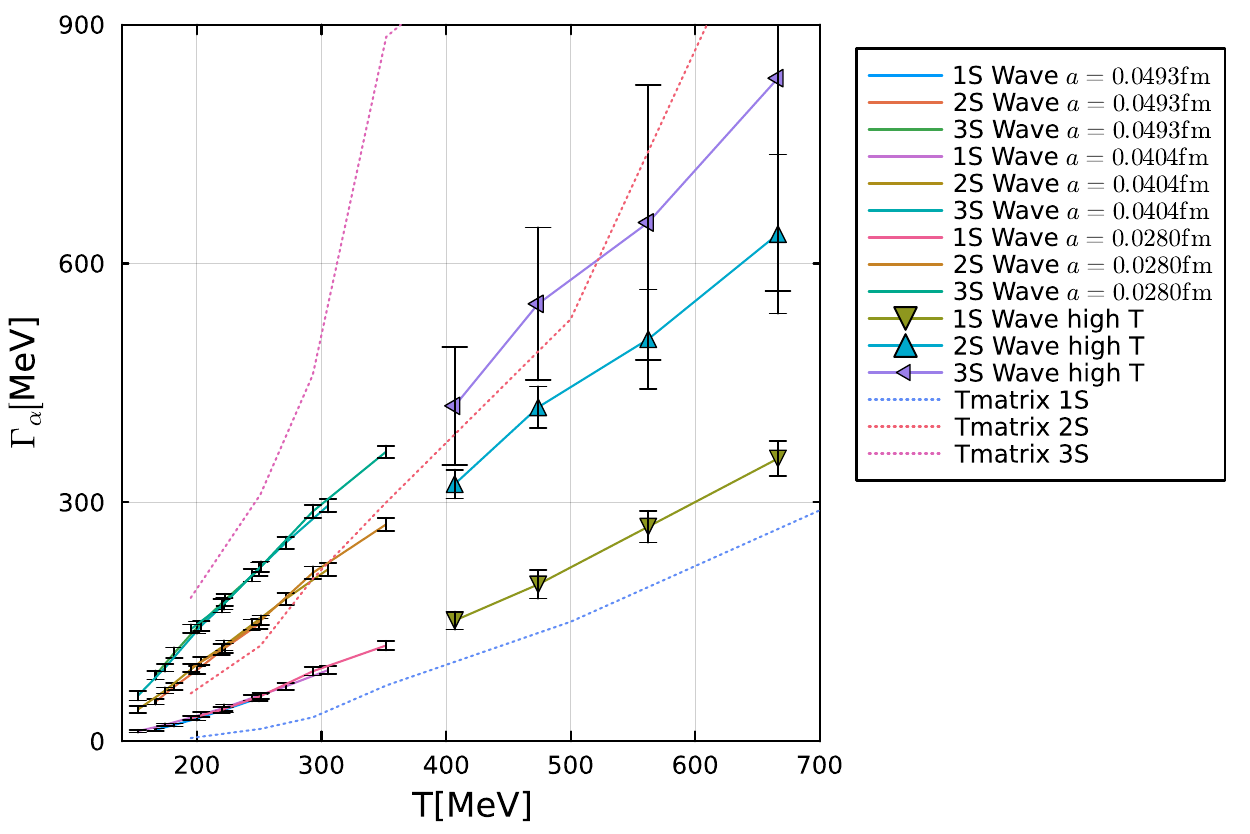}
\includegraphics[width=0.48\textwidth]{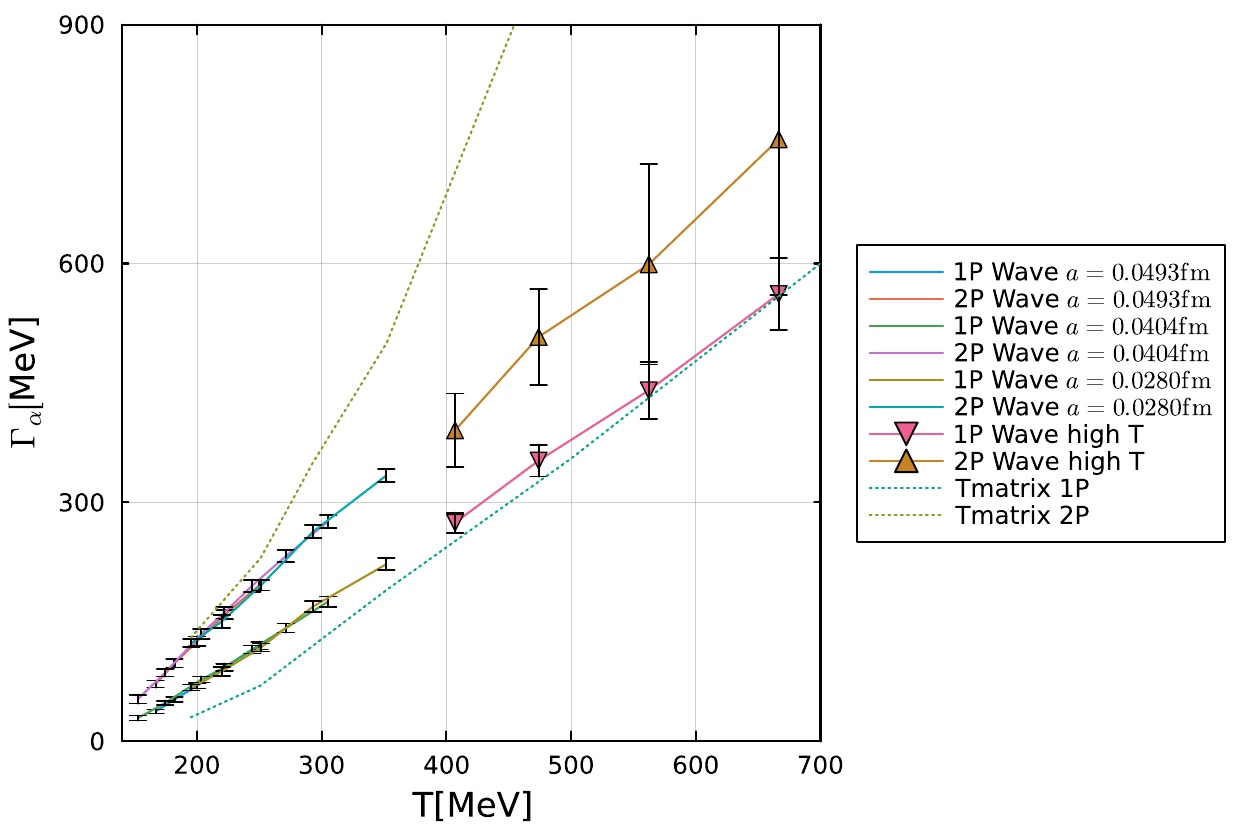}
\includegraphics[width=0.48\textwidth]{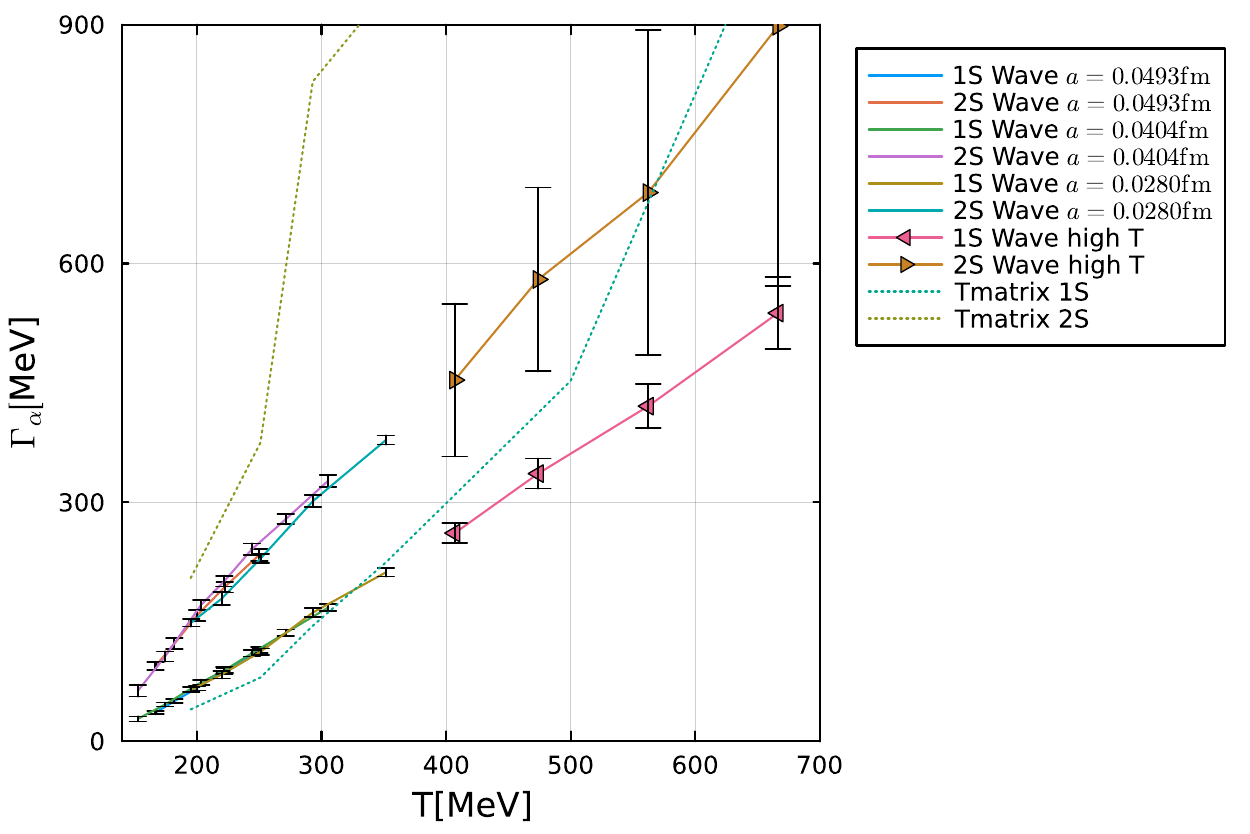}
\includegraphics[width=0.48\textwidth]{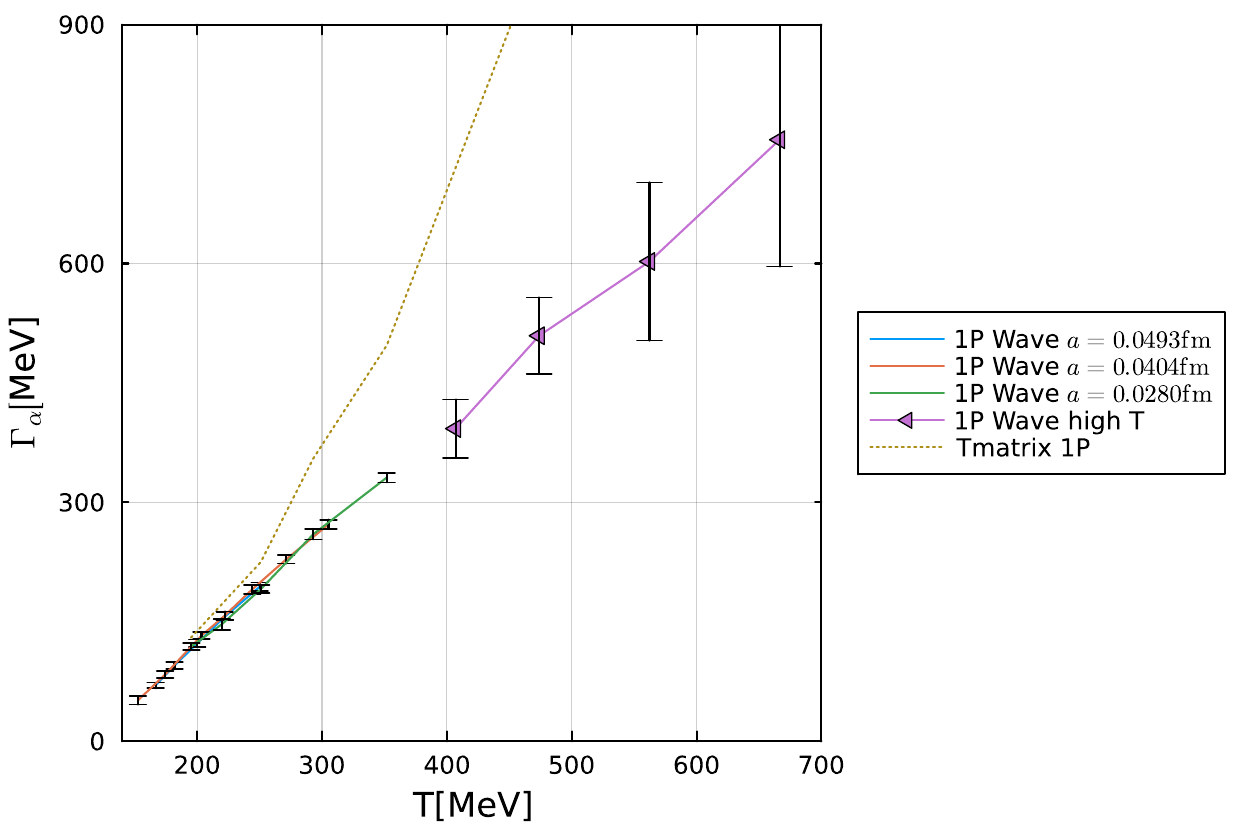}
\caption{Comparison of the spatial-average estimates (Wave) with the $T$-matrix results of Ref.~\cite{Wu:2025hlf}: $S$-wave bottomonia (top left), $P$-wave bottomonia (top right), $S$-wave charmonia (bottom left), and $P$-wave charmonia (bottom right). 
The wave functions and treatment of the imaginary potential are described in the text. The high T estimates are from small volumes of size $(48a)^3$ which only contains wilson line correlator results up to $r/a=15$. For distances larger than that, we do an upper and lower estimate, assuming the data continues linearly at larger separations, or that it stays constant. The results shown for high T are a combination of statistical errors and half the difference in the estimates, while the shown value is the average.}
\label{fig:T-matrix}
\end{figure}

Having assessed the model against direct lattice estimates, we use the imaginary potential to extend the quarkonium width estimates to higher temperatures, up to  $700$~MeV. We compare these results with the $T$-matrix calculations of Ref.~\cite{Wu:2025hlf}, whose interaction kernel is constrained by comparisons with lattice Wilson line and bottomonium correlators in the temperature range $T=195$--$352$~MeV. Outside this range, the calculation is not directly constrained by these correlator data. The comparison is shown in Fig.~\ref{fig:T-matrix}.
The differences between our estimates and the $T$-matrix results can be understood in terms of the spectral tails, which have been cut in Eq.~\eqref{eq:pot_peak}, and color screening. For relatively narrow states, the tails contribute appreciably to the Euclidean correlator, and including them allows a smaller peak width to reproduce the same correlator, as demonstrated in the lattice NRQCD analysis~\cite{Ding:2025fvo}. This can explain the smaller $T$-matrix widths for $1S$ quarkonia and $1P$ bottomonia at lower temperatures. As the widths increase, the sensitivity to the tails diminishes, as the shape of the spectral peak around the peak position
plays a more prominent role in determining the details of the $\tau$ dependence of the Euclidean correlations
functions. For this reason our approach and the T-matrix calculations yield similar results.
For the $1S$ and $1P$ bottomonium states, the small screening effects present in
the T-matrix approach does not play an important role.
For higher excited states, however, screening becomes important. Although the screening of the interaction kernel in the $T$-matrix approach is considerably weaker than in earlier studies, it still reduces the binding and promotes a rapid increase of the widths above a state-dependent temperature~\cite{Wu:2025hlf}. Our model, which uses vacuum wave functions motivated by the absence of appreciable screening in the lattice potential, does not include this additional effect and therefore predicts a more gradual increase of the excited-state widths.

\section{Conclusions}
\label{sec:conclusions}

In this paper we revisited the determination of the complex $Q \bar Q$ potential from lattice QCD. In particular, using lattice QCD simulations on larger spatial volumes, we studied the finite volume effects on the Wilson line correlator. In the vacuum  the finite volume effects are the most pronounced at small Euclidean time, while at larger Euclidean time these are modest or non-existent. 
This is the relevant region for the extraction of the complex potential, and although residual systematic effects cannot be excluded, the analysis indicates that the medium-dependent contributions are comparatively insensitive to the spatial volume within the present precision. Furthermore, the spectral analysis on the larger volumes yields results consistent with those obtained previously on smaller volumes, supporting their use in determining the complex potential.

We revisited the determination of the imaginary part of the potential based on the previous lattice QCD data using a more realistic Ansatz for the dominant peak in the spectral function, which resulted in a smaller imaginary part of the potential. Using these new numerical results on the imaginary part of  the potential, we pointed out that there is a connection between the imaginary part of the potential and the quarkonium width given by Eq. \eqref{eq:width} and validated it through comparison to the  thermal width of different quarkonium states obtained from charmonium \cite{Larsen:2026qvs} and bottomonium \cite{Ding:2025fvo} correlation functions. 

Then we use this relation to predict the width of different quarkonium states for temperatures where no lattice results are available yet and only model calculations based on the $T$-matrix approach exist.  For temperatures below 300 MeV our width estimate for $1S$ and $1P$ bottomonia are larger than the $T$-matrix result, which is likely due to simple modeling of the spectral peak. At higher temperatures, we see agreement between the $T$-matrix and our results for 1P bottomonium state.  There are larger differences between our estimates and the $T$-matrix results for $2P$ and $3S$ bottomonia as well as for $2S$ charmonia for $T>250$ MeV with our calculations giving significantly smaller results for the thermal width due to a lack of significant color screening effects.

\section*{Acknowledgments}

This material is based on work supported by the U.S. Department of Energy, Office of Science, Office of Nuclear Physics under Contract No. DE-SC0012704  and the Topical Collaboration in Nuclear Theory "Heavy-Flavor Theory (HEFTY) for QCD Matter".

R.N.L. was supported by the Ministry of Culture and Science of the State of Northrhine Westphalia (MKW NRW) under the funding code NW21-024-A (NRW-FAIR). R.N.L. acknowledge support from the Deutsche Forschungsgemeinschaft (DFG)
through the CRC-TR 211 “Strong-interaction matter under extreme conditions” (Project No.
315477589 - TRR 211).

This research used computing time provided by the ALCC and INCITE programs at the Oak Ridge Leadership Computing Facility, a DOE Office of Science User Facility supported under contract no. DE-AC05-00OR22725; the National Energy Research Scientific Computing Center (NERSC), a DOE Office of Science User Facility at Lawrence Berkeley National Laboratory under Contract No. DE-AC02-05CH11231; PRACE awards on JUWELS at GCS@FZJ, Germany; Marconi100 at CINECA, Italy; and EuroHPC JU award for the project ID EHPC-EXT-2024E01-039 access to the LUMI-G supercomputer at the Finnish IT Center for Science (CSC). Part of the computations in this work were performed on the GPU cluster at Bielefeld University. We thank the Bielefeld HPC.NRW team for their support.

\appendix
\section{Appendix: Lattice setup and finite-volume effects}
\label{sec:setup}

\begin{table}[t]
\centering
\begin{tabular}{ccccc}
\toprule
$N_s$ & $N_\tau$ & $T$ [MeV] & \# conf. & $t_F/a^2$ \\
\midrule
\colorrow
64 & 16 & 250 & 4697 & $0.0,\,0.2,\,0.4,\,0.6$ \\
\colorrow
64 & 18 & 222 & 3715 & $0.0,\,0.2,\,0.4,\,0.6$ \\
\colorrow
64 & 20 & 200 & 3005 & $0.0,\,0.2,\,0.4,\,0.6$ \\
\colorrow
64 & 22 & 182 & 4158 & $0.0,\,0.2,\,0.4,\,0.6$ \\
\colorrow
64 & 24 & 167 & 3278 & $0.0,\,0.2,\,0.4,\,0.6$ \\
\colorrow
64 & 26 & 154 & 2423 & $0.0,\,0.2,\,0.4,\,0.6$ \\
\midrule
96 & 16 & 250 & 8800 & $0.0,\,0.2,\,0.4,\,0.6$ \\
96 & 18 & 222 & 7200 & $0.0,\,0.2,\,0.4,\,0.6$ \\
96 & 20 & 200 & 5800 & $0.0,\,0.2,\,0.4,\,0.6$ \\
96 & 22 & 182 & 4600 & $0.0,\,0.2,\,0.4,\,0.6$ \\
96 & 24 & 167 & 3600 & $0.0,\,0.2,\,0.4,\,0.6$ \\
96 & 26 & 154 & 2900 & $0.0,\,0.2,\,0.4,\,0.6$ \\
\midrule
\colorrow
64 & 64 & 63 & 914   & $0.0,\,0.2,\,0.4,\,0.6$ \\
80 & 80 & 50 & 3199  & $0.0,\,0.2,\,0.4,\,0.6$ \\
\bottomrule
\end{tabular}
\caption{Lattice ensembles used for the finite-volume analysis at
$\beta=7.596$, corresponding to $a=0.0493~{\rm fm}$. The pink highlighted rows correspond to ensembles taken from previous works, while the others were recently generated. }
\label{tab:volume_ensembles}
\end{table}

\begin{table}[t]
\centering
\begin{tabular}{ccccc}
\toprule
$N_s$ & $N_\tau$ & $T$ [MeV] & \# conf. & $t_F/a^2$ \\
\midrule
\colorrow
64 & 16 & 305 & 5528 & $0.0,\,0.2,\,0.4,\,0.6$ \\
\colorrow
64 & 18 & 271 & 5230 & $0.0,\,0.2,\,0.4,\,0.6$ \\
\colorrow
64 & 20 & 244 & 4726 & $0.0,\,0.2,\,0.4,\,0.6$ \\
\colorrow
64 & 22 & 222 & 3515 & $0.0,\,0.2,\,0.4,\,0.6$ \\
\colorrow
64 & 24 & 203 & 3345 & $0.0,\,0.2,\,0.4,\,0.6$ \\
\midrule
128 & 16 & 305 & 1176 & $0.0,\,0.2,\,0.4,\,0.6$ \\
128 & 18 & 271 &  822 & $0.0,\,0.2,\,0.4,\,0.6$ \\
128 & 20 & 244 &  780 & $0.0,\,0.2,\,0.4,\,0.6$ \\
128 & 22 & 222 &  700 & $0.0,\,0.2,\,0.4,\,0.6$ \\
128 & 24 & 203 &  625 & $0.0,\,0.2,\,0.4,\,0.6$ \\
\midrule
\colorrow
64 & 64 & 76 & 1006 & $0.0,\,0.2,\,0.4,\,0.6$ \\
96 & 96 & 51 & 2074 & $0.0,\,0.2,\,0.4,\,0.6$ \\
\bottomrule
\end{tabular}
\caption{Lattice ensembles used for the finite-volume analysis at
$\beta=7.825$, corresponding to $a=0.0404~{\rm fm}$. The pink highlighted rows correspond to ensembles taken from previous works, while the others were recently generated.}
\label{tab:volume_ensembles_b7825}
\end{table}

In this appendix, we provide additional numerical details of the Wilson line correlator (WLC) calculations and investigate the sensitivity of the results to the spatial lattice volume.

We computed WLC in (2+1)-flavor QCD using the highly improved staggered quark (HISQ) action~\cite{Follana:2006rc} for the fermions and the L{\"u}scher-Weisz gauge action~\cite{Weisz:1982zw,Luscher:1984xn}. To investigate finite-volume effects, we extend the calculation to larger spatial volumes at two values of the bare gauge coupling. The ensembles used in this analysis are summarized in Tables~\ref{tab:volume_ensembles} and~\ref{tab:volume_ensembles_b7825}. 
The lattice spacing was determined using the $r_1$ scale.

At $\beta=7.596$, corresponding to $a=0.0493~{\rm fm}$, we compare the $64^4$ and $80^4$ zero-temperature ensembles, with spatial extents of approximately $3.16$ and $3.94~{\rm fm}$, respectively. At finite temperature, the $64^3\times N_\tau$ ensembles are compared with new $96^3\times N_\tau$ ensembles at the same bare parameters and temporal extents. The spatial extent is therefore increased from approximately $3.16~{\rm fm}$ to $4.73~{\rm fm}$ while keeping the lattice spacing and temperature unchanged.

A second finite-volume comparison is performed at $\beta=7.825$, corresponding to $a=0.0404~{\rm fm}$. In this case, the zero-temperature reference ensembles have volumes $64^4$ and $96^4$, corresponding to spatial extents of approximately $2.59$ and $3.88~{\rm fm}$. The finite-temperature comparison involves $64^3\times N_\tau$ ensembles and new $128^3\times N_\tau$ ensembles, corresponding to spatial extents of $2.59$ and $5.17$ fm, respectively.

These two sets of ensembles allow us to study the spatial-volume dependence both directly in the finite-temperature correlators and through the zero-temperature reference correlators entering the subtraction procedure.

On each gauge configuration we calculate temporal WLC in Coulomb gauge. To improve the signal at large Euclidean separations, we apply gradient flow (GF) to the gauge fields prior to gauge fixing. More specifically, we 
smooth the gauge fields over a characteristic radius $r_F=\sqrt{8\tau_F}$, where $\tau_F$ denotes the flow time. Increasing the flow time substantially reduces the statistical fluctuations of the Wilson-line correlators, particularly at large $\tau$, but also modifies the correlator at distances comparable to the flow radius. We therefore consider several values of $\tau_F$ and restrict the physical interpretation of the data to Euclidean separations for which these short-distance flow effects are under control. After the flow evolution, the configurations are fixed to Coulomb gauge with a gauge-fixing tolerance of $10^{-6}$.

We further reduce statistical fluctuations by exploiting the smooth dependence of the correlators on the spatial separation \(r\). At fixed Euclidean time, we perform a local quadratic interpolation using correlators evaluated at nearby separations. This procedure improves the statistical precision of the interpolated correlators, although neighboring interpolated values are correlated because their interpolation windows may overlap.

The WLC requires 
an additive normalization of the effective mass, Eq. \eqref{eq:meff_cont}. This shift is determined from the zero-temperature static potential and depends on the flow time because gradient flow modifies the ultraviolet self-energy contribution. Effective masses obtained at different flow times are therefore aligned through a flow-time-dependent additive shift determined from the zero-temperature data in a region where flow-induced short-distance distortions are negligible.

For each separation, the zero-temperature ground-state contribution is determined by fitting the correlator in a suitable Euclidean-time interval to Eq.~\eqref{eq:WLC_lowest_state}. The resulting overlap $A_r$ and static potential $V(r,0)$ are used to construct the high-energy contribution through~Eq. \eqref{eq:WLC_high}, which is subsequently subtracted from the finite-temperature correlator. The subtraction is performed independently using the corresponding small- and large-volume zero-temperature ensembles. 

Statistical uncertainties, including those associated with the determination of the zero-temperature ground-state contribution, are propagated through the subtraction using a jackknife analysis. Since the zero- and finite-temperature correlators are obtained from statistically independent ensembles, their uncertainties are evaluated separately. The finite-temperature contribution is determined by varying the finite-temperature correlator over its jackknife samples while keeping the zero-temperature input fixed. Independently, for each zero-temperature jackknife sample, the ground-state parameters $A_r$ and $V(r,0)$ are fitted and the corresponding high-energy contribution $W^{\rm high}(\tau,r)$ is reconstructed per sample before performing the subtraction. This preserves the correlations between the zero-temperature correlator and its fitted ground-state contribution. The resulting finite- and zero-temperature jackknife variances are then added in quadrature to obtain the statistical uncertainty of the subtracted observables.

\begin{figure}[t]
    \centering
    \includegraphics[page=4,width=0.45\linewidth]{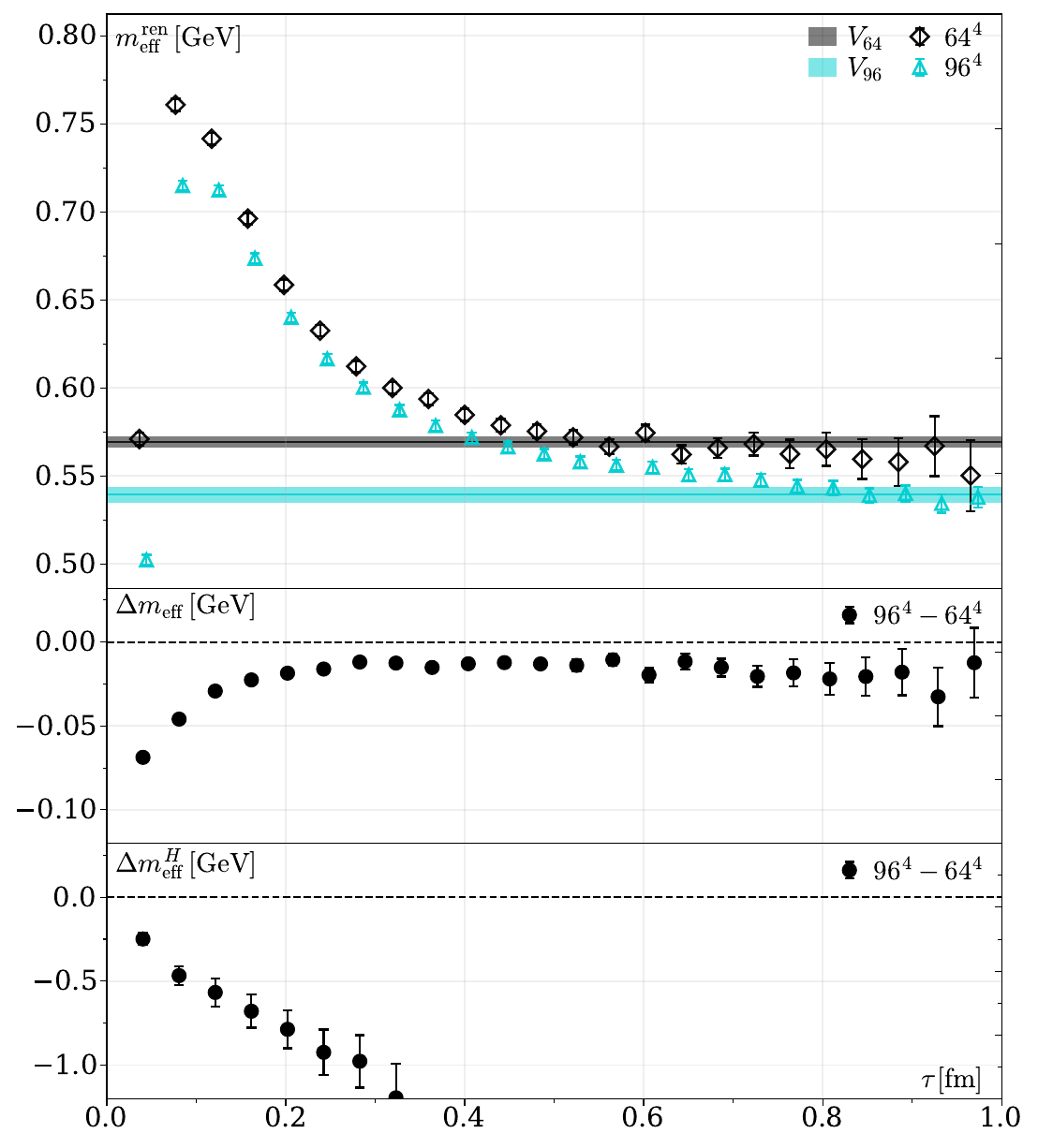}
    \caption{Zero-temperature finite-volume comparison at $\beta=7.825$, $r/a=15$ $(r\simeq0.61$ fm), and $\tau_F/a^2=0.6$. The upper panel shows the renormalized effective masses on the $64^4$ and $96^4$ ensembles, together with the corresponding static-potential estimates. The middle panel shows the difference between the effective masses on the two volumes, while the lower panel shows the corresponding difference for the residual high-energy contribution defined in Eq.~\eqref{eq:meff_high}.}
    \label{fig:zerot-finite-volume}
\end{figure}

\begin{figure*}[t]
    \centering
    \begin{subfigure}[t]{0.45\textwidth}
        \centering
        \includegraphics[page=4,width=\linewidth]{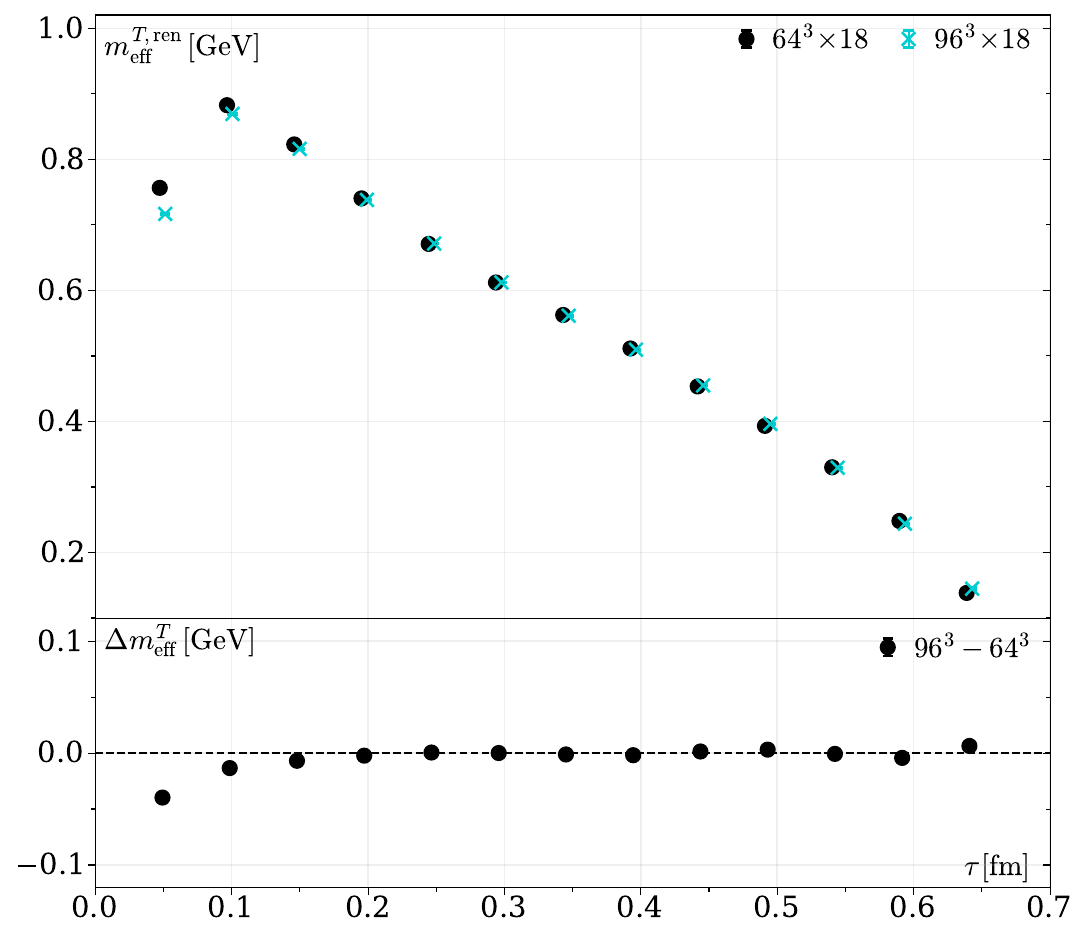}
        \caption{$\beta=7.596$ and $T=222$ MeV}
    \end{subfigure}
    \begin{subfigure}[t]{0.45\textwidth}
        \centering
        \includegraphics[page=4,width=\linewidth]{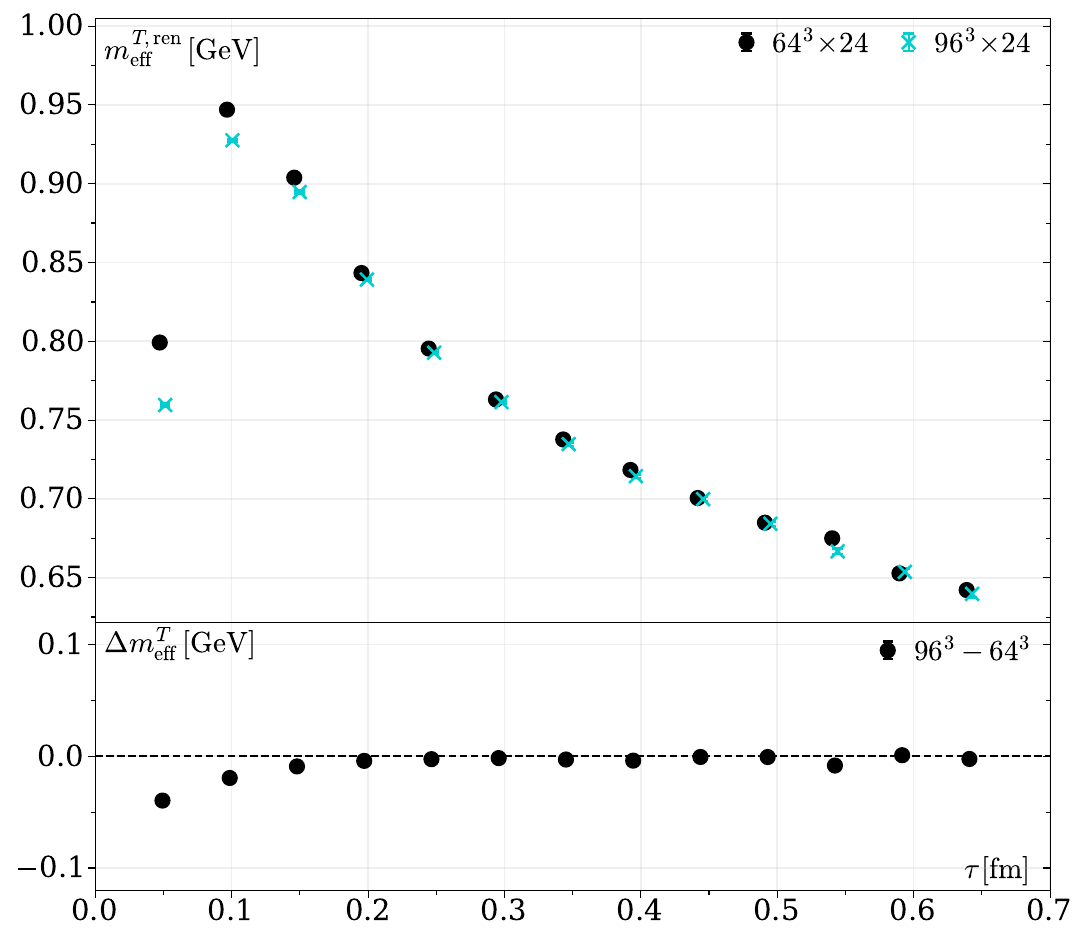}
        \caption{$\beta=7.596$ and $T=167$ MeV}
    \end{subfigure}
    \begin{subfigure}[t]{0.45\textwidth}
        \centering
        \includegraphics[page=4,width=\linewidth]{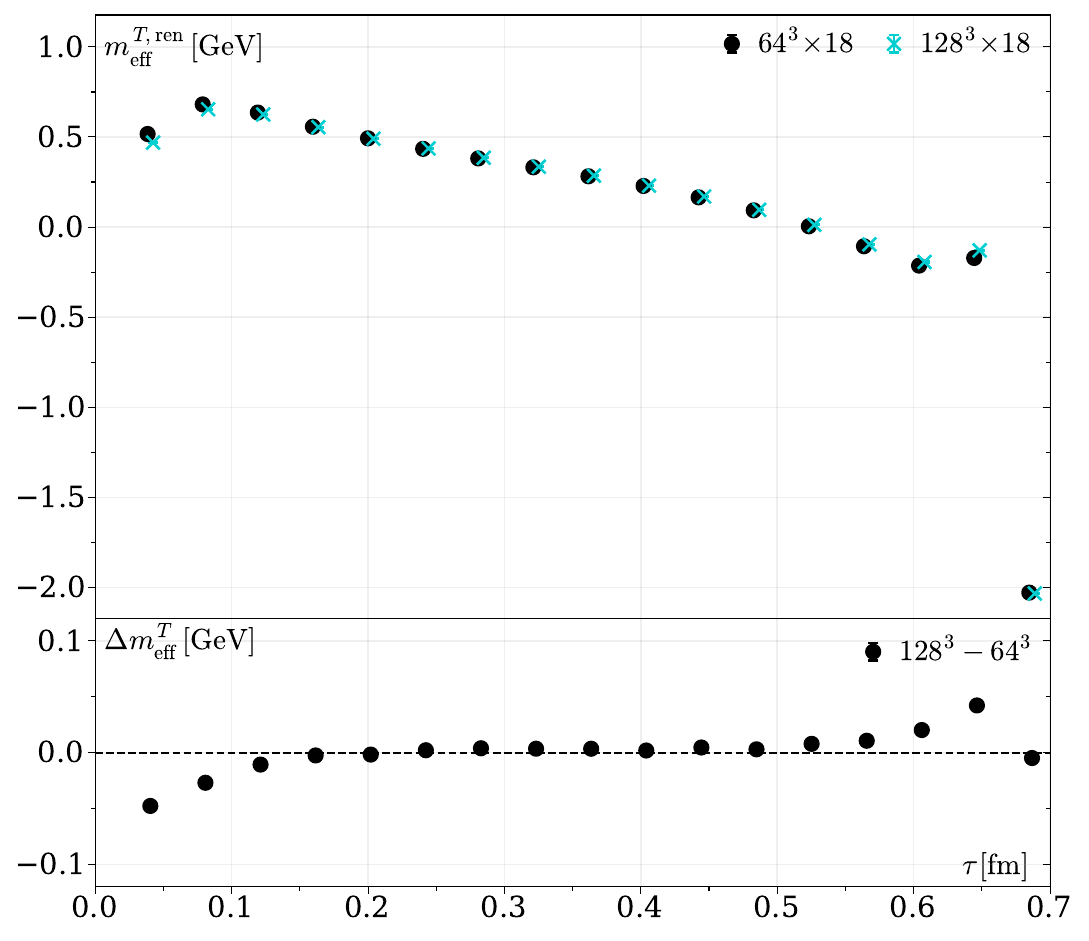}
        \caption{$\beta=7.825$ and $T=271$ MeV}
    \end{subfigure}
    \begin{subfigure}[t]{0.45\textwidth}
        \centering
        \includegraphics[page=4,width=\linewidth]{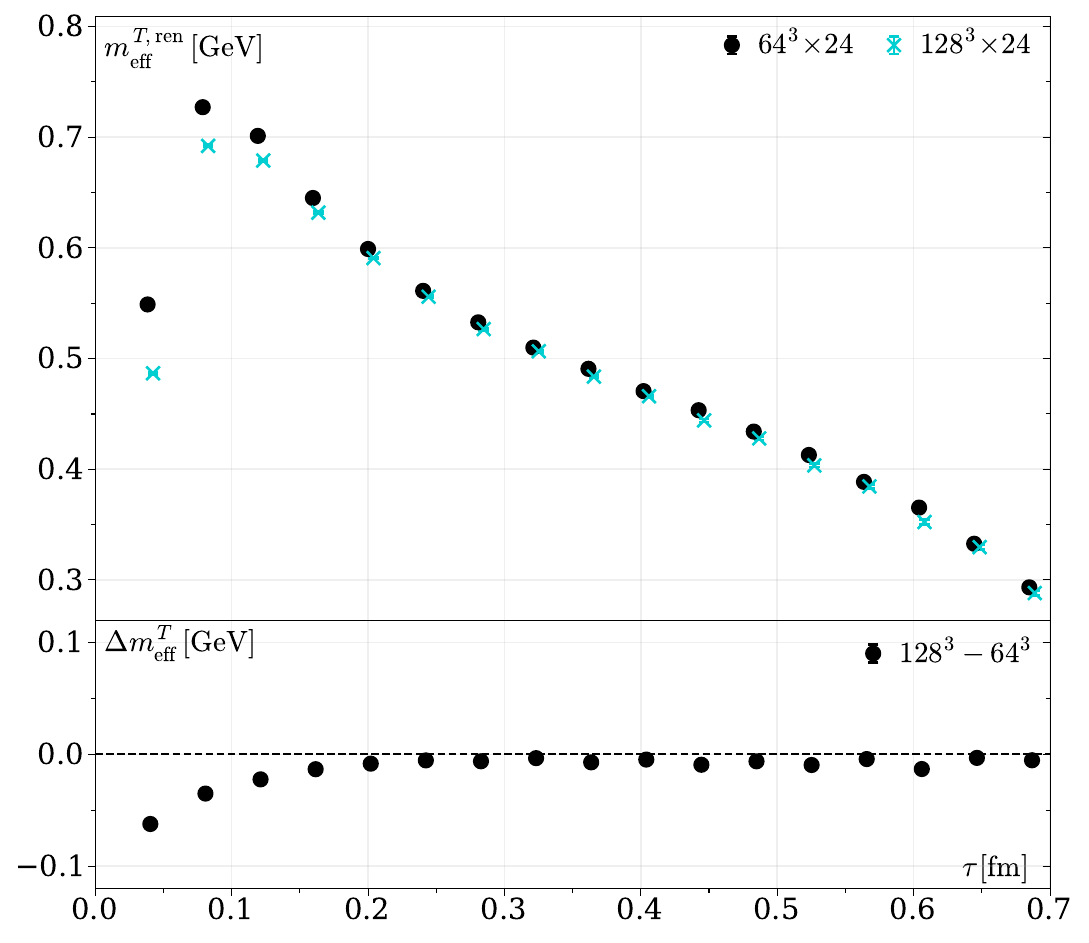}
        \caption{$\beta=7.825$ and $T=203$ MeV}
    \end{subfigure}
    
    \caption{Finite-volume comparison of the unsubtracted finite-temperature effective masses at $\tau_F/a^2=0.6$ and $r/a=15$. The upper panels correspond to $\beta=7.596$, with spatial volumes $64^3$ and $96^3$, at $T=222$ MeV (left) and $167$ MeV (right). The lower panels show the corresponding comparison at $\beta=7.825$, with spatial volumes $64^3$ and $128^3$, at $T=271$ MeV (left) and $203$ MeV (right). The lower panel of each subplot shows the difference between the effective masses on the larger and smaller volumes.}
    \label{fig:finite-volume-direct-temperature-dependence}
\end{figure*}

\begin{figure*}[t]
    \centering
    \begin{subfigure}[t]{0.45\textwidth}
        \centering
        \includegraphics[width=\linewidth]{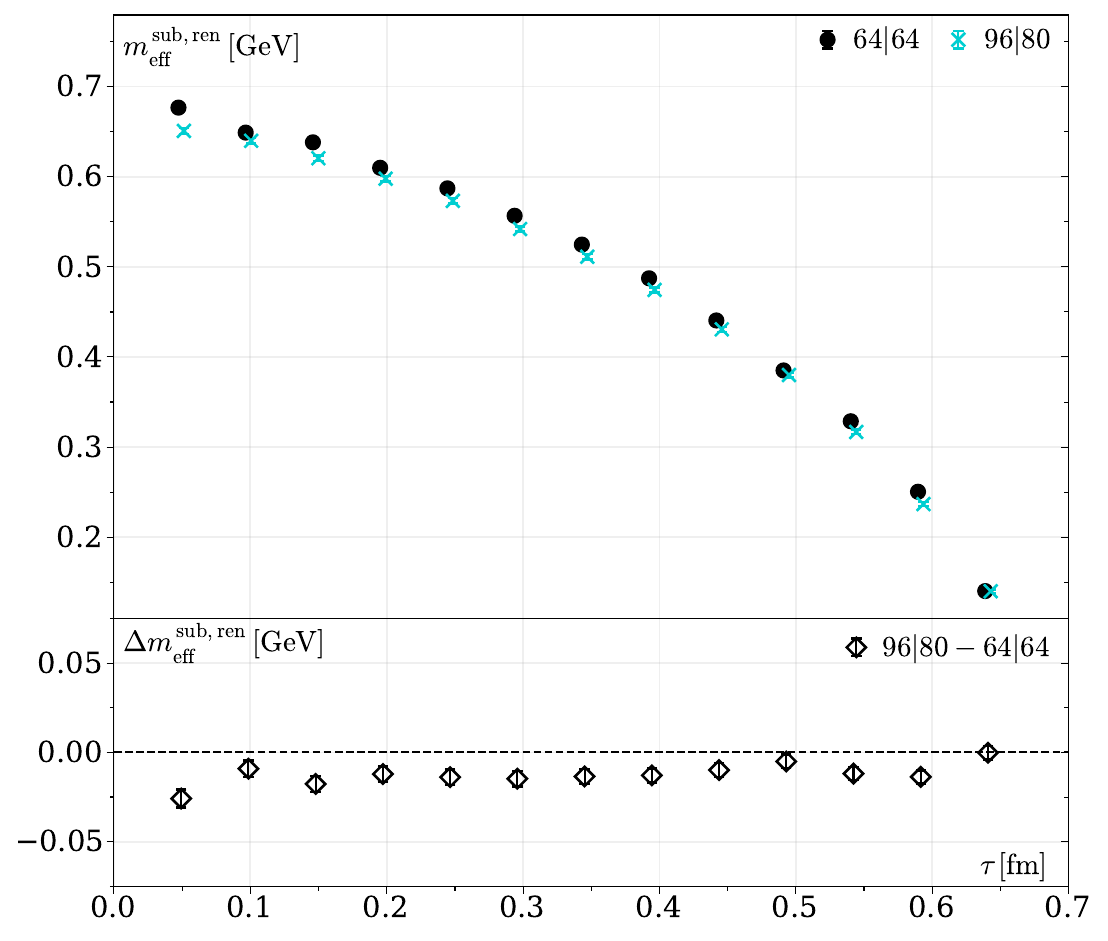}
        \caption{$\beta=7.596$ and $T=222$ MeV}
    \end{subfigure}
    \begin{subfigure}[t]{0.45\textwidth}
        \centering
        \includegraphics[width=\linewidth]{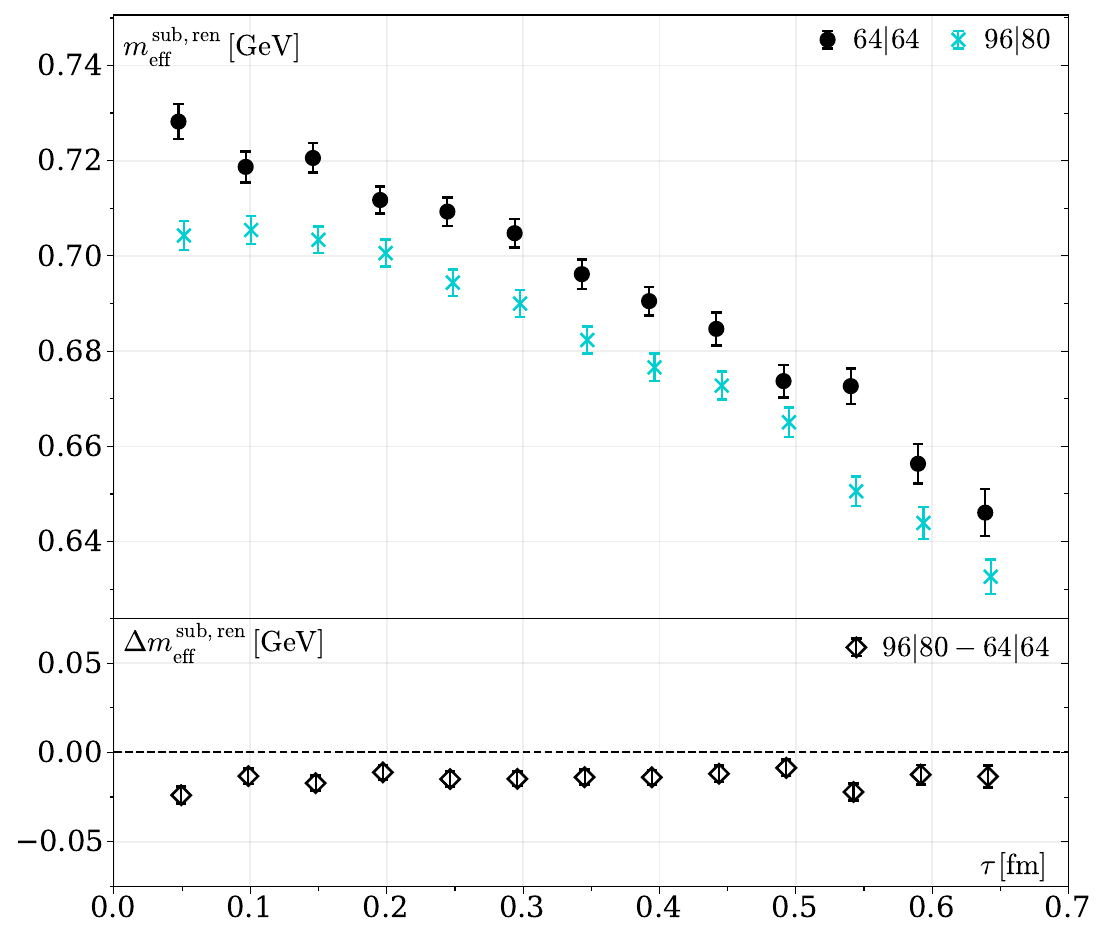}
        \caption{$\beta=7.596$ and $T=167$ MeV}
    \end{subfigure}
    \begin{subfigure}[t]{0.45\textwidth}
        \centering
        \includegraphics[width=\linewidth]{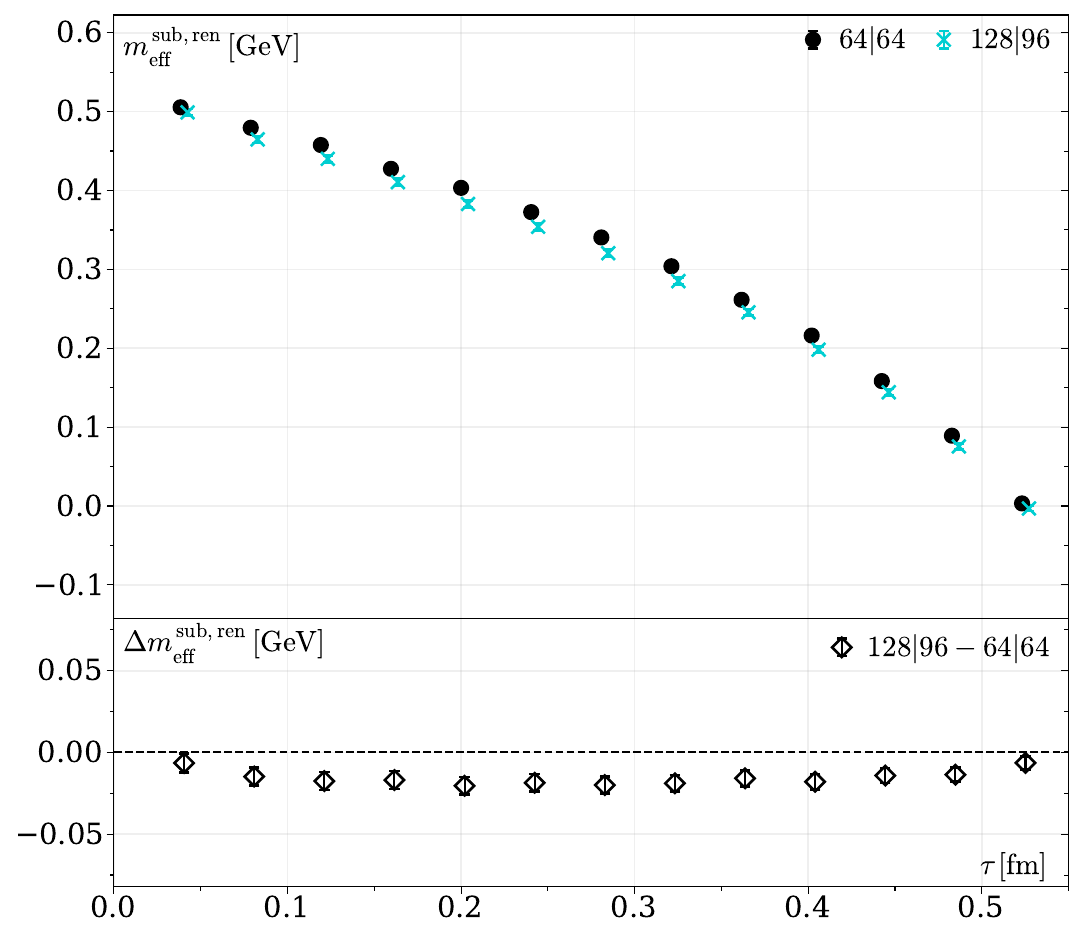}
        \caption{$\beta=7.825$ and $T=271$ MeV}
    \end{subfigure}
    \begin{subfigure}[t]{0.45\textwidth}
        \centering
        \includegraphics[width=\linewidth]{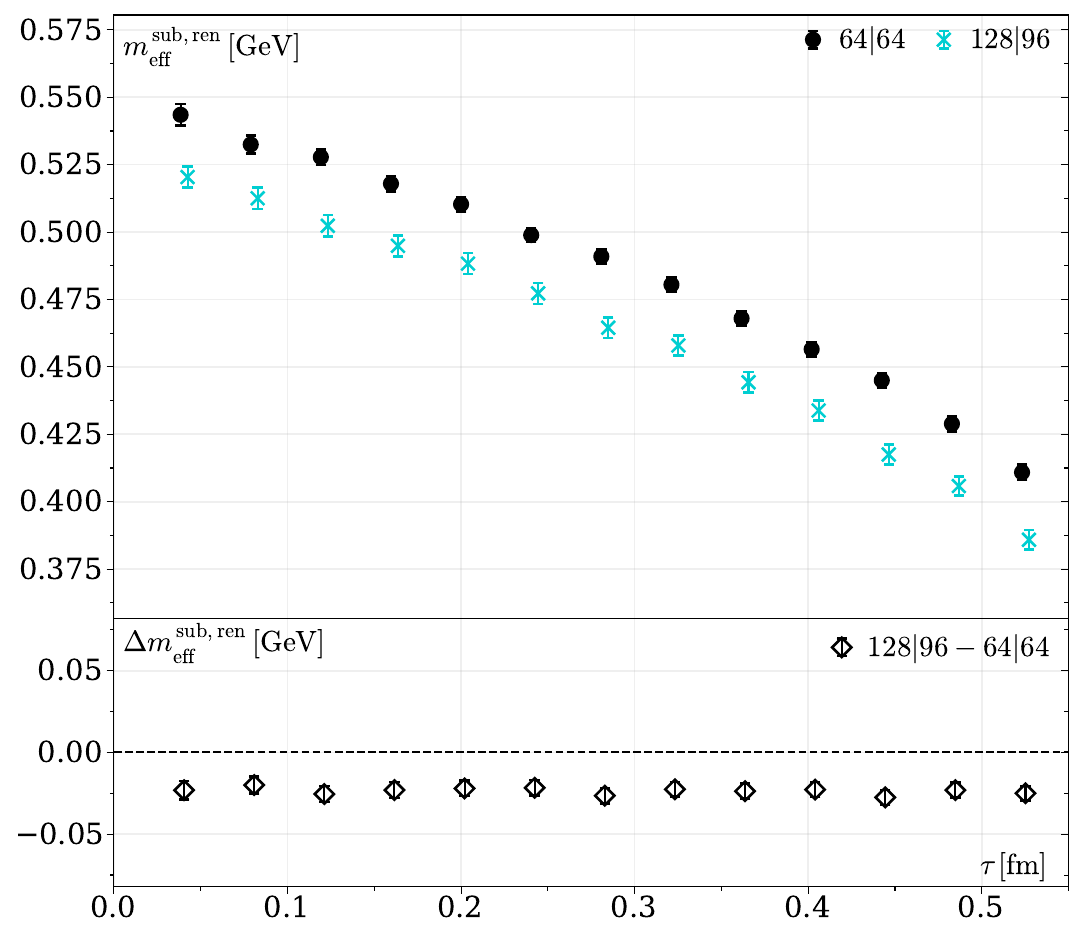}
        \caption{$\beta=7.825$ and $T=203$ MeV}
    \end{subfigure}
    
    \caption{Finite-volume comparison of the subtracted effective masses at the same lattice parameters and temperatures as in Fig.~\ref{fig:finite-volume-direct-temperature-dependence}. The upper panels compare the $64|64$ and $96|80$ combinations at $\beta=7.596$, while the lower panels compare the $64|64$ and $128|96$ combinations at $\beta=7.825$. The notation $L|L_0$ specifies the spatial extents of the finite- and zero-temperature ensembles used in the subtraction. The lower panel of each subplot shows the difference between the corresponding large- and small-volume results.}
    \label{fig:finite-volume-temperature-dependence}
\end{figure*}

We now turn to the finite-volume comparison. We focus on the representative lattice separation $r/a=15$, corresponding to $r\simeq0.74$ fm at $\beta=7.596$ and $r\simeq0.61$ fm at $\beta=7.825$. We have also examined other separations to assess the spatial dependence of the observed effects.

Finite-volume dependence can enter the analysis in two ways: through the finite-temperature correlator itself and through the zero-temperature reference used to construct the subtracted correlator. We examine these contributions separately before considering their combined effect on the subtracted effective masses.

Figure \ref{fig:meff_T0} shows the zero-temperature comparison at $\beta=7.596$, while Fig.~\ref{fig:zerot-finite-volume} presents the corresponding results at $\beta=7.825$. In both cases, the renormalized effective masses exhibit a visible dependence on the spatial volume, particularly at smaller Euclidean-time separations. The volume dependence is more pronounced in the effective masses associated with the residual high-energy contribution, $m_{\mathrm{eff}}^{\mathrm{high}}$, after subtraction of the fitted ground state.

This behavior suggests that a substantial part of the observed finite-volume dependence originates from the excited-state and higher-energy contributions to the correlator. However, the comparison also indicates a possible volume dependence of the extracted ground-state contribution, particularly at the finer lattice spacing. This can be seen directly in the upper panel bands of Figs.~\ref{fig:meff_T0} and \ref{fig:zerot-finite-volume}. Since the determination of the static potential relies on identifying a ground-state-dominated Euclidean-time region, the stability of this extraction with respect to the fit interval is an important consideration.

We next examine the finite-temperature effective masses before applying the zero-temperature subtraction. Figure \ref{fig:finite-volume-direct-temperature-dependence} compares the results obtained on different spatial volumes at both lattice spacings. At $\beta=7.596$, the upper panels show representative temperatures $T=222$ and $167$ MeV, comparing $64^3\times N_\tau$ and $96^3\times N_\tau$ ensembles. At $\beta=7.825$, the lower panels present analogous comparisons at $T=271$ and $203$ MeV, using $64^3\times N_\tau$ and $128^3\times N_\tau$ ensembles.

The effective masses generally exhibit good agreement between the two spatial volumes over the Euclidean-time intervals where the statistical uncertainties remain under control. Nevertheless, some volume dependence is visible, especially at small $\tau$, with differences depending on the temperature and Euclidean-time separation. 

Finally, we investigate the finite-volume dependence of the subtracted effective masses. Following Sec.~\ref{sec:complex}, we use the notation $N_s^T|N_s^0$, where $N_s$ and $N_s^0$ denote the spatial extents, in lattice units, of the finite- and zero-temperature ensembles entering the subtraction. At $\beta=7.596$, the default small- and large-volume combinations are $64|64$ and $96|80$, respectively. At $\beta=7.825$, the corresponding combinations are $64|64$ and $128|96$.

Figure \ref{fig:finite-volume-temperature-dependence} shows the resulting subtracted effective masses at the same temperatures considered in Fig.~\ref{fig:finite-volume-direct-temperature-dependence}. The comparisons indicate that increasing the spatial volume modifies the subtracted effective masses only moderately over the Euclidean-time region relevant for the extraction of the complex potential, giving an overall, almost constant effect of around 20 MeV. The exact size and Euclidean-time dependence of these differences vary between the two lattice spacings and temperatures.

Because the zero- and finite-temperature spatial extents are changed simultaneously in these comparisons, the observed differences incorporate both the direct volume dependence of the thermal correlator and the sensitivity to the zero-temperature reference. The agreement after subtraction should therefore be interpreted together with the separate comparisons of the original correlators discussed above. 

Overall, the finite-volume analysis indicates that the higher-energy part of the zero-temperature correlator exhibits a more pronounced spatial-volume dependence than the original finite-temperature effective masses. The subtracted observables are comparatively stable over the Euclidean-time region used in the analysis, although residual volume dependence, including possible effects in the determination of the zero-temperature static potential, cannot be excluded. These results provide an assessment of the finite-volume systematic effects entering the extraction of the complex potential.

We finally note that while the spatial volumes of the zero- and finite-temperature ensembles are identical for the small-volume combinations, this is not the case for the larger-volume comparisons. In particular, the $96|80$ and $128|96$ combinations employ zero-temperature reference correlators calculated on smaller spatial volumes than their finite-temperature counterparts. Consequently, part of the observed differences in the subtracted effective masses may arise from this volume mismatch rather than from the intrinsic finite-volume dependence of the thermal correlators. The present comparison does not allow these contributions to be disentangled unambiguously, and this represents an additional (possibly small) source of systematic uncertainty in the subtraction procedure.

\bibliographystyle{apsrev4-1}
\bibliography{ref}

\end{document}